\documentclass{aastex63}

\usepackage{float}
\usepackage{svg}
\usepackage{upgreek}
\usepackage{lipsum}
\usepackage[version=4]{mhchem}

\received{November 1, 2024}
\revised{April 8, 2025}
\accepted{May 14th, 2025}
\submitjournal{ApJ}

\shorttitle{Modern Earth Chemical Disequilibrium Biosignature Constraints}
\shortauthors{Young et al.}
\graphicspath{{./}{figures/}}

\begin{document}


\title{Modern Earth-like Chemical Disequilibrium Biosignatures Are Challenging To Constrain Through Spectroscopic Retrievals}

\correspondingauthor{Amber Young}
\email{amber.v.young@nasa.gov, amberastro12@gmail.com}

\author[0000-0003-3099-1506]{Amber V. Young}
\affiliation{NASA Goddard Space Flight Center \\
8800 Greenbelt Rd. \\
Greenbelt, Maryland 20770, USA}
\affiliation{Department of Astronomy and Planetary Science, Northern Arizona University, Flagstaff, Arizona 86011, USA}

\author[0000-0002-3196-414X]{Tyler D. Robinson}
\affiliation{Lunar \& Planetary Laboratory, University of Arizona, Tucson, AZ 85721 USA}
\affiliation{Department of Astronomy and Planetary Science, Northern Arizona University, Flagstaff, Arizona 86011, USA}
\affiliation{Habitability, Atmospheres, and Biosignatures Laboratory, University of Arizona, Tucson, AZ 85721, USA}
\affiliation{NASA Nexus for Exoplanet System Science Virtual Planetary Laboratory, University of Washington, Box 351580, Seattle, WA 98195, USA}


\author{Joshua Krissansen-Totton}
\affiliation{Department of Earth and Space Sciences/Astrobiology Program, University of Washington, Seattle WA 98195}

\author[0000-0002-2949-2163]{Edward W. Schwieterman}
\affiliation{Department of Earth and Planetary Sciences, University of California, Riverside, CA, USA}
\affiliation{Blue Marble Space Institute of Science, Seattle, WA, 98154, USA}

\author{Giada Arney}
\affiliation{NASA Goddard Space Flight Center \\
8800 Greenbelt Rd. \\
Greenbelt, Maryland 20770, USA}

\author[0000-0002-5292-4200]{Gerrick E. Lindberg}
\affiliation{Department of Chemistry and Biochemistry, Northern Arizona University, Flagstaff, Arizona 86011, USA}

\author{Cristina Thomas}
\affiliation{Department of Astronomy and Planetary Science, Northern Arizona University, Flagstaff, Arizona 86011, USA}



\begin{abstract}
Robust exoplanet characterization studies are underway, and the community is looking ahead toward developing observational strategies to search for life beyond our solar system. With the development of life detection approaches like searching for atmospheric chemical species indicative of life, chemical disequilibrium has also been proposed as a potentially key signature for life. Chemical disequilibrium can arise from the production of waste gases due to biological processes and can be quantified using a metric known as the available Gibbs free energy. The main goal of this study was to explore the detectability of chemical disequilibrium for a modern Earth-like analog. Atmospheric retrievals coupled to a thermodynamics model were used to determine posterior distributions for the available Gibbs free energy given simulated observations at various noise levels. In reflected light, chemical disequilibrium signals were difficult to detect and limited by the constraints on the \ce{CH4} abundance, which was challenging to constrain for a modern Earth case with simulated observations spanning ultraviolet through near-infrared wavelengths with V-band SNRs of 10, 20, and 40. For a modern Earth analog orbiting a late-type M dwarf, we simulated transit observations with the \textit{James Webb Space Telescope} Mid-Infrared Instrument (MIRI) and found that tight constraints on the available Gibbs free energy can be achieved, but only at extremely low noise on the order of several ppm. This study serves as further proof of concept for remotely inferring chemical disequilibrium biosignatures and should be included in continuing to build life detection strategies for future exoplanet characterization missions.
\end{abstract}

\keywords{Astrobiology, Exoplanet Atmospheres, Habitable Planets}


\section{Introduction} \label{sec:intro}
The exoplanet science community is entering a new era of exoplanet characterization and biosignature assessment. Mission efforts like NASA's \textit{Kepler} \citep{Borucki_2010_Kepler} and \textit{Transiting Exoplanet Survey Satellite} (\textit{TESS}) \citep{Ricker_2015_TESS} have revealed thousands of exoplanets, many of them Earth-sized making potentially Earth-like habitable exoplanets rather common \citep{Petigura_2013,Dressing_2013,Kriso_Charbonneau_2023}. As more observational data is collected, now is an important time to develop novel techniques and metrics for interpreting exoplanet atmospheric signals that may be indicative of life. We have made great strides in determining the signals of life (i.e., biosignatures) that we would be able to search for on exoplanets using remote spectroscopic observations, including the detection of specific atmospheric species \citep[e.g.,][]{Schwieterman_2018}. For example, O$_2$, has been studied extensively because of its direct ties to biological processes like photosynthesis, which is globally exhibited on modern Earth \citep{Meadows_2017,Meadows_2018}. 

Future exoplanet characterization missions such as NASA's future Habitable Worlds Observatory will benefit from continued development of prior biosignature detection strategies. One potential exoplanet biosignature that has seen relatively little study\,---\,chemical disequilibrium\,---\,can arise from the atmospheric accumulation of waste gases produced by metabolic processes. The substantial production of these waste gases can then significantly influence the atmosphere and its thermochemical state. One such example in Earth's atmosphere is the co-existence of molecular oxygen (\ce{O2}) and methane (\ce{CH4}). In this chemical context, large biological fluxes of \ce{CH4} are needed to maintain its continuous presence in the highly oxidized environment of the atmosphere \citep{Lovelock1975,Sagan1993,Simoncini_2013,krissansen-totton_detecting_2016}. 

The inferred connection between chemical disequilibrium and biological processes has been posed as a fundamental indicator of life since the work of \citet{lovelock_physical_1965} \citep[see also][]{Hitchcock1967, Lovelock1975}. It was thought that life could substantially influence the chemical composition of its biosphere similarly to how life on Earth has shaped the chemical composition of the atmosphere throughout geologic history. Chemical disequilibrium biosignatures are agnostic in the sense that they are not specific to a given metabolism, only that the atmosphere is perturbed out of chemical equilibrium because of that metabolism. Since then, a metric for quantifying chemical disequilibrium has been derived and termed the available Gibbs free energy \citep{Lovelock1975,krissansen-totton_detecting_2016,Krissansen-Totton_2018}. The Gibbs free energy is a thermodynamic state function that describes the maximum amount of work that can be done by a given process, which is minimized at equilibrium \citep{engelthomas_thermodynamics_2019}. \citet{krissansen-totton_detecting_2016} demonstrated that the extent of chemical disequilibrium can be quantified by taking the difference between the observed Gibbs free energy of an atmospheric (or atmosphere-ocean) system and the theoretical equilibrium Gibbs free energy for that same system, which is deemed the ``available Gibbs free energy". In an earlier study, the available Gibbs free energy of a simulated Proterozic Earth-like planet was found to be detectable at a high abundance scenario and constrained to within an order of magnitude with SNR 50 observations \citep{AVYoung_2024}. The material below now carefully describes the coupling techniques that were previously developed in order to pair remote observations to a thermodynamics model and we investigate how challenging is it to remotely infer the available Gibbs free energy for a modern Earth-like exoplanet. 

Thermodynamics modeling\,---\,that can tell us about the available Gibbs free energy of a given planetary atmosphere\,---\, combined with spectral observations and analysis techniques can fill the knowledge gap for how chemical disequilibrium biosignatures can be remotely inferred in practice. Spectral observations can allow us to interpret various information about an exoplanet's atmospheric state, including properties such as gas mixing ratios, global surface pressure, effective temperature, and physical properties of other opacity sources (e.g., clouds, aerosols, and hazes). This is a challenging endeavour as the information has to be extracted from noisy spectra. Nevertheless, retrieval analysis techniques can be used to extract this information from planetary spectra \citep[e.g.,][]{Madhusudhan_2009,Benneke_2012,Line_2013,Lupu_2016,Feng_2018,Barstow_2020,MacDonald_2023} and put constraints on the parameters needed to calculate the available Gibbs free energy. Methods for calculating the chemical disequilibrium of an Earth system were coupled to simulated observations and retrieval analyses in order to constrain the available Gibbs free energy of a modern Earth-like exoplanet analog both in reflected light and in transit. This allows us to test our ability to remotely detect and quantify chemical disequilibrium signatures and to evaluate how available Gibbs free energy constraints are sensitive to observational uncertainty. These analyses are important for establishing which observational constraints are most relevant for constraining fundamental chemical disequilibrium biosignatures and for further building life detection strategies for future exoplanet missions.

\section{Methods} \label{sec:meth}
The material that follows describes the adopted atmospheric retrieval and thermochemical tools. After these overviews, a coupling procedure is presented that, then, enables a remote sensing approach to quantifying constraints on the available Gibbs free energy.

\subsection{Atmospheric Retrieval Model}
The \texttt{rfast} retrieval model incorporates a radiative transfer forward model, an instrument noise simulator, and a Bayesian statistical analysis tool in order to investigate the atmospheric state of a given planetary atmosphere \citep{Robinson_2023}. \texttt{rfast} can perform simulated observations in both 1D and 3D for a given planetary scenario and can simulate reflected light, emission, and transit spectra. We simulated 1D reflected light and transit spectral observations for a modern Earth-like exoplanet analog. In reflected light, the inhomogeneous atmospheric reflectivity is calculated recursively via an adding approach, with,
\begin{equation}
    R^{+}_{j,N} = r_j + \frac{t^{2}_{j}R^{+}_{j+1,N}}{1-r_{j}R^{+}_{j+1,N}} \ ,
    \label{Reflc.EQ}
\end{equation}
where $R^{+}_{j,N}$ is defined here as the reflectance of a column of atmosphere from the bottom of the atmosphere ($j=N$) to an atmospheric layer ($j$; i.e., adding upward), $r_j$ represents the reflectivity of a given atmospheric layer ($j$), and $t_j$ is the transmissivity of this layer. By computing the optical properties due to scattering and absorption of each layer and recursively computing $R^{+}_{j,N}$, we arrive at a total atmospheric reflectivity, which is akin to geometric albedo ($A_{\rm g}$). This wavelength dependent geometric albedo is then used to model the planet-to-star flux ratio,
\begin{equation}
\frac{F_{\rm p}}{F_{\rm s}} = A_{\rm g} \phi(\alpha) \left(\frac{R_{\rm p}}{a}\right)^2 \ ,
\label{Fp_Fs_EQ}
\end{equation}
where $A_{\rm g}$ is the geometric albedo, $\phi$ is the phase function (a function of observational phase angle $\alpha$, and set equal to unity in the 1D approach), $R_{\rm p}$ is the radius of the planet, and $a$ is the orbital distance.  

For \texttt{rfast} transit spectroscopy, the wavelength dependent spectrum is derived from  a one-dimensional (radial) vectorized approach \citep{Robinson_2023,Robinson_2017}, with,
\begin{equation}
    \left(\frac{R_{\rm p,\lambda}}{R_{\rm s}}\right)^2 = \frac{1}{R{\rm s}^2}\left(R_{\rm p}^2 + \frac{1}{\pi}\textbf{a}_{\lambda} \cdot \textbf{A}\right)
\end{equation}
where $R_{\rm p}$ is the planetary radius, $R_{\rm s}$ is the stellar radius, $\textbf{a}_{\lambda}$ is wavelength-dependent absorptivity, and \textbf{A} a vector of annulus areas for the planetary atmosphere silhouetted on the stellar disk.

The instrument noise model apart of \texttt{rfast} was used to simulate constant signal-to-noise along the full wavelength range. This generalized approach helps mitigate instrument assumptions which is particularly relevant for the direct imaging scenarios where the instrumentation has yet to be finalized for HWO. The high resolution simulated spectra from the forward model are degraded to the user specified instrument resolution. Finally, the retrieval framework for \texttt{rfast} uses Markov Chain Monte Carlo (MCMC) analysis implemented with the open source python tool \texttt{emcee} \citep{Foreman_Mackey_2013}. The MCMC analysis entails examining the range of potential values for each retrieval parameter by utilizing prior probabilities and the likelihood of the observed data, given a specific instance of the model fit. The resultant posterior probability distributions are representative of the regions of parameter space that are most likely to fit the observed data (whether it be real or simulated data generated with the noise model). For each atmospheric scenario, 10 independent MCMC chains were generated, with each chain using a unique instance of the data spectrum simulated using randomized spectral errors. Each chain consists of 200 walkers and 100,000 steps. To exclude non-converged regions of the parameter space, a ``burn-in" period of 50,000 steps is removed from each chain. The post-burn-in segments of the 10 chains are then randomly sampled and combined into a final chain, which integrates the randomized effects from the individual data spectra. This approach follows established practices in \citet{Feng_2018}. To validate this setup, sensitivity tests were conducted on an example case by doubling the number of steps to 200,000. These tests showed negligible differences in the resulting marginal posterior distributions, providing strong evidence that 100,000 steps are sufficient to achieve convergence. 

\subsection{Gibbs Free Energy Model}
The Gibbs free energy model we use originated from \citet{krissansen-totton_detecting_2016} and is a tool used to calculate the extent of chemical disequilibrium in a given system quantified by the available Gibbs free energy. The total temperature- and pressure-dependent Gibbs free energy of formation for the gas phase is calculated by:
\begin{equation}
    \Delta_{\rm f} G(T,P,\{N_i\}) = \sum_{i} \left[\Delta_{\rm f}G_{i}^{\circ}(T,P^{\circ}) + R_{\rm U}T\ln{\left(\frac{N_i P}{N P^{\circ}}\gamma_{fi}\right)} \right] N_i \ ,
\end{equation}
where $T$ is temperature, $P$ is pressure, $\Delta_{\rm f}G_{i}^{\circ} (T,P_{\rm r})$ is the temperature-dependent Gibbs free energy of formation for species `$i$' at standard state pressure ($P^{\circ}$), $R_{\rm U}$ is the Universal gas constant, $N_i$ is the number of moles for a given species, $N$ is the total number of moles, and $\gamma_{f,i}$ is the activity coefficient for a given species. The sum is over all included species. Because the Gibbs free energy is minimized for a chemical system in equilibrium at constant pressure and temperature, we can assess the theoretical equilibrium state of the system by finding chemical species abundances ($N_i$) required to minimize the above expression using non-linear optimization. The difference in Gibbs free energy between the initial (or observed) state and the final (or equilibrium) state allows us to quantify how the observed state deviates from its theoretical equilibrium state, thereby yielding a metric for chemical disequilibrium termed the available Gibbs free energy,
\begin{equation}
    \Phi \equiv \Delta_{\rm f} G(T,P,\{N_i\}_{\rm obs}) - \Delta_{\rm f} G(T,P,\{N_i\}_{\rm eqm}).
\end{equation}
The Gibbs free energies computed herein are all gas phase calculations. Our focus on gas phase Gibbs free energy calculations is due to exoplanet spectroscopy being particularly sensitive to atmospheric/planetary properties. Inferences of oceanic chemical disequilibrium on the other hand may require in-situ observations and gathering knowledge of oceanic parameters (e.g., dissolved species and pH), which is beyond current exoplanet observational capabilities.

\subsubsection{A Modern Earth Analog Case Study}
\begin{figure}[H]
    \centering
    \includegraphics[scale=0.35]{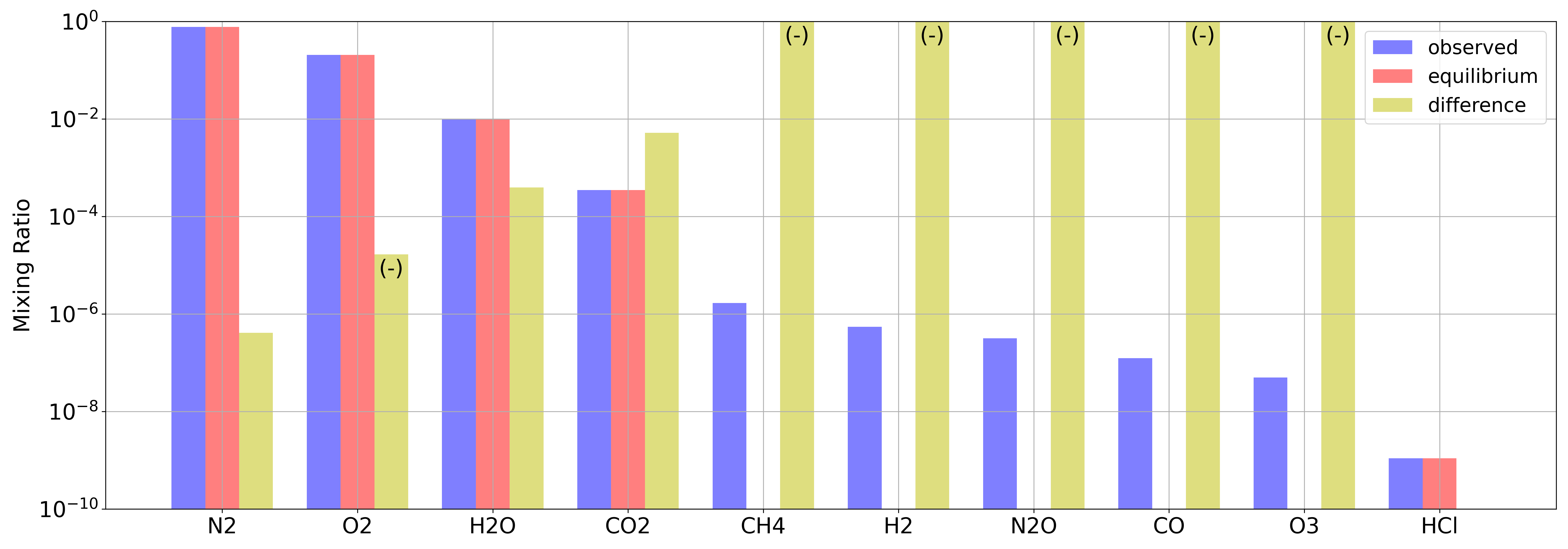}
    \caption{A modern Earth thermodynamic calculation using the Gibbs free energy thermodynamics model. Each species included in the calculation is listed along the bottom with their respective mixing ratios on the y-axis. The blue bars represent the observed abundances for each species and the red bars indicate the equilibrium abundance of each species. The yellow bars show the fractional difference in between the observed and equilibrium abundances, normalized by the observed values. The ``(-)" annotations indicate a decrease in mixing ratio takes place in order to reach equilibrium. This calculation was done at a pressure of 1\,bar and a  temperature of 288\,K.}
    \label{ME_Gibbs_Calc}
\end{figure}

Figure \ref{ME_Gibbs_Calc} is the gas phase chemical disequilibrium calculation of a modern Earth twin using the Gibbs free energy model. The x-axis lists the chemical species included in the calculation (i.e., \ce{N2}, \ce{O2}, \ce{H2O}, \ce{Ar}, \ce{CO2}, \ce{Ne}, \ce{He}, \ce{CH4}, \ce{Kr}, \ce{H2}, \ce{N2O}, \ce{CO}, \ce{Xe}, \ce{O3}, and \ce{HCl}) with the exception of the inert species, which have abundances that do not vary between the observed and equilibrium states. The blue vertical bars represent the observed abundances (from \textit{in situ} data here), the red bars indicate the equilibrium abundances, and the yellow bars indicate the fractional difference in abundance between the observed and equilibrium values (normalized by the observed values) for each chemical species. The available free energy for this atmospheric scenario is 1.51\,J\,mol$^{-1}$. 
\noindent Chemical species that exhibit a difference between their observed and equilibrium abundance are in disequilibrium. We narrowed down the list of retrieved chemical species (outlined in Table \ref{tab_retrieval}) 
for coupling the thermodynamics model to the retrievals. The atmospheric gases with the most influence were in disequilibrium and had a high relative abundance in comparison to the other species.

\subsection{Coupled Model}
To couple atmospheric inferences with thermodynamic computations of available Gibbs free energy, we randomly sample the marginal posterior distributions of relevant retrieved parameters (listed in bold in Table \ref{tab_retrieval}) and pass them as inputs to the Gibbs free energy model. These parameters include surface pressure ($P_0$), characteristic atmospheric temperature ($T_0$), and chemical species mixing ratios of \ce{O2}, \ce{H2O}, \ce{CO2}, \ce{O3}, and \ce{CH4}. Upon specifying the abundance for each species, the atmosphere is then back-filled with \ce{N2} as the background gas.  Repeating the random sampling and coupling process thousands of times produced a resulting marginal posterior distribution for the available Gibbs free energy. The atmospheric gases included in the retrieval were chosen because they exhibited the most influence (by far) on the magnitude of the disequilibrium due to their reactivity and/or high abundance compared to neglected species (e.g., \ce{N2O} and \ce{H2}).

\begin{figure}[H]
    \centering
    \includegraphics[scale=0.5]{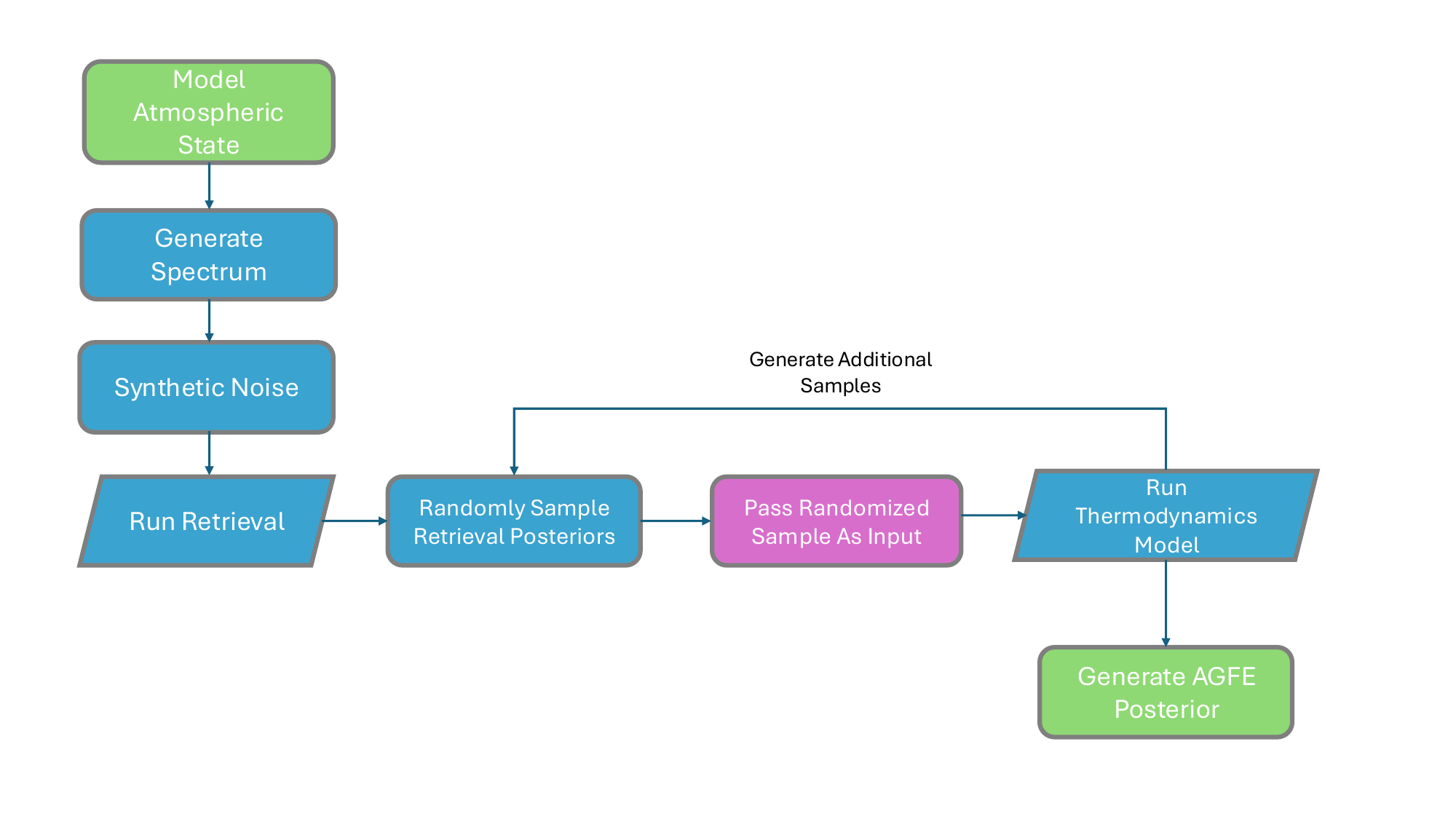}
    \caption{A schematic that outlines the step-by-step process for coupling the atmospheric retrievals to the thermodynamics model. In green are the start and end steps, which include starting with an initial atmospheric state and ending with a statistical posterior for the available Gibbs free energy (AGFE). The purple step indicates the point at which the retrievals are coupled to the thermodynamics modeling via random sampling of the retrieval parameter posteriors. The process of randomly sampling the retrieval posteriors and running the thermodynamics model are repeated until the desired number of samples is reached, allowing the generation of the AGFE posterior.}
    \label{schematic}
\end{figure}

\begin{table}[h]
    \centering
    \begin{tabular}{l c c c}
    \hline
    \hline
         Parameter & Description & Prior  \\
         \hline
         \textbf{log} $\mathbf P_0$ (log Pa) & Surface pressure & [0,8]\\
         $\mathbf T_0$ (K) & Atmospheric temperature & [100,1000]\\
         \textbf{log O$_2$} & Molecular oxygen mixing ratio & [-10,0]\\
         \textbf{log H$_2$O} & Water vapor mixing ratio & [-10,0]\\
         \textbf{log CO$_2$} & Carbon dioxide mixing ratio& [-10,0]\\
         \textbf{log O$_3$} & Ozone mixing ratio&[-10,-2]\\
         \textbf{log CH$_4$} & Methane mixing ratio&[-10,0]\\
         log $A\rm_s$ & Surface albedo&[-2,0]\\
         log $R\rm_p$ (log R$_\oplus$) & Planetary Radius &[-1,1]\\
         log $M\rm_p$ (log M$_\oplus$) & Planetary Mass&[-1,2]\\
         log $\Delta p \rm_c$ (log Pa) & Cloud thickness &[0,8]\\
         log $p\rm_t$ (log Pa) & Cloud top pressure&[0,8]\\
         log $\tau\rm_c$ & Cloud optical depth&[-3,3]\\
         log $f\rm_c$ & Cloud fraction&[-3,0]\\
         \hline
         \hline
    \end{tabular}
    \caption{List of 14 planetary parameters that were retrieved in this work using  \texttt{rfast}. Listed are the parameter names, descriptions, and their corresponding priors (which are flat across the indicated range). The \textbf{bold} parameters were randomly sampled to compute the posterior distributions for the available Gibbs free energy. Clouds were not included in the transit retrievals, which thereby exclude log $\Delta$p$\rm_c$, log p$\rm_t$, log $\tau\rm_c$, and log f$\rm_c$ for those specific cases. Additionally, within the context of the transit retrievals, M$\rm_p$ and R$\rm_p$ were retrieved in linear space with analogous prior ranges spanning 0.1\,--\,10 Earth masses and Earth radii, respectively.}
    \label{tab_retrieval}
\end{table}

\section{Results} \label{sec:res}
The Gibbs free energy calculation relies on species abundance information as well as planetary surface pressure and characteristic atmospheric temperature, which are all key to modeling the thermodynamic state. To make Gibbs free energy inferences, a forward model for a specified planetary atmospheric state is run, and synthetic noise is added to the resulting spectrum. Then a retrieval model is applied to the faux observation, yielding information about how the faux observation constrains the atmospheric state. The outputs of the retrieval are then randomly sampled and passed as inputs to the thermodynamics model. Each thermodynamic calculation is modeled as a closed system of the atmospheric state, which is characterized by the randomly drawn volume mixing ratios of the gas phase species, global surface pressure, and characteristic atmospheric temperature.  

\subsection{The Planetary Spectra}
\begin{figure}[H]
    \centering
    \includegraphics[scale=0.6]{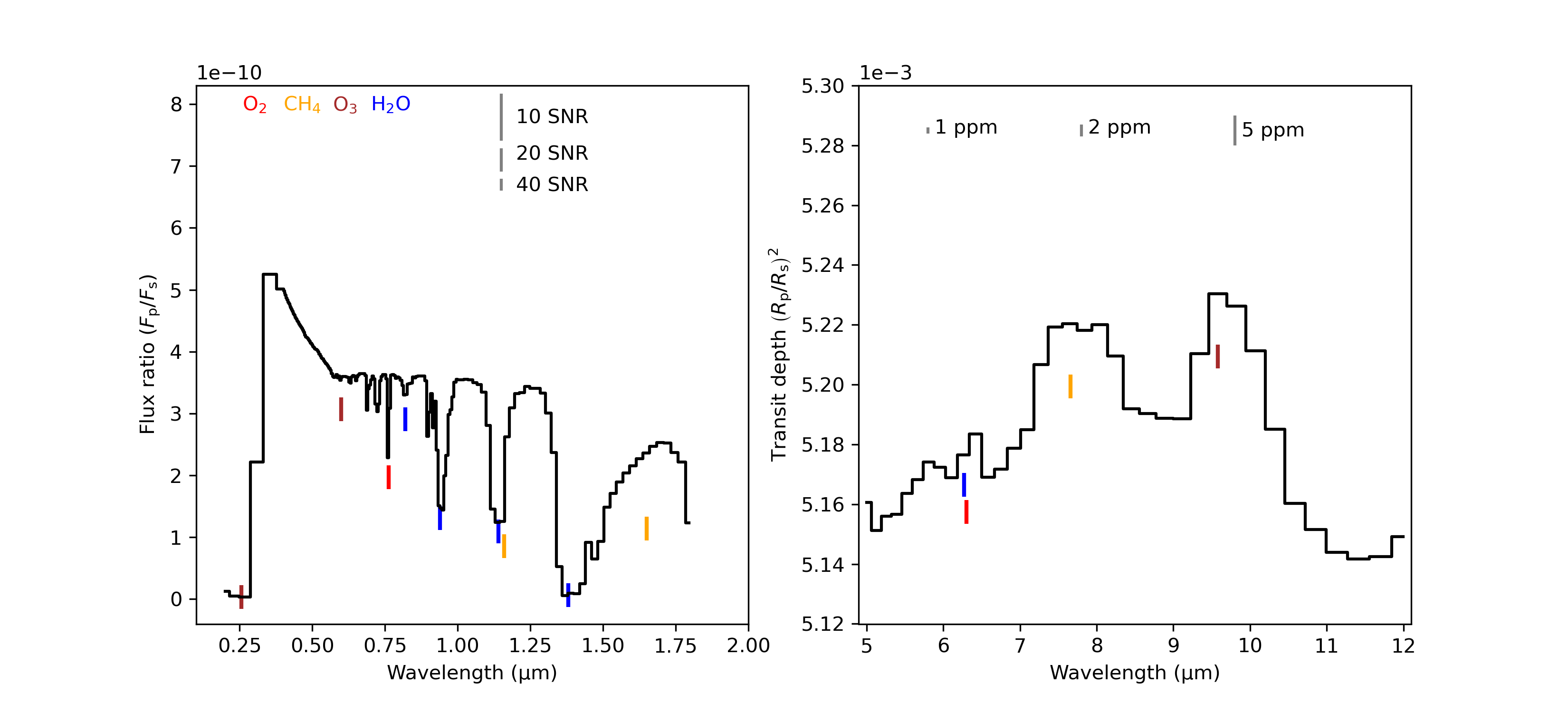}
    \caption{Spectra for modern Earth in reflected light (left) and a modern Earth analog atmospheric setup for TRAPPIST-1e in transit (right). The color coded tick marks indicate the central wavelength of the absorption feature for a given species where blue ticks indicate \ce{H2O} features, brown ticks are for \ce{O3} features, orange ticks are for \ce{CH4} features, and red ticks indicate \ce{O2} absorption features. The scale of the error bars for each observational scenario are indicated in gray next to each case.}
    \label{spectra}
\end{figure}

Figure \ref{spectra} shows a simulated Ultraviolet\,--\,visible\,--\,Near-infrared reflected light spectrum (left) and simulated transit spectrum (right) with various species absorption features labeled in color coded tick marks. Earth's atmospheric chemical disequilibrium is primarily maintained through the co-existence of \ce{O2} and \ce{CH4} \citep{krissansen-totton_detecting_2016,Krissansen-Totton_2018}. In reflected light, the abundance constraints for \ce{O2} are mainly obtained through the strong absorption of the \ce{O2} A-band at 0.76\,$\upmu$m. The \ce{CH4} abundance is best constrained by its absorption in the near infrared at 1.16\,$\upmu$m and 1.65\,$\upmu$m. However, \ce{CH4} abundance inferences are made challenging by the \ce{H2O} feature at 1.14\,$\upmu$m which blends with \ce{CH4} and the overall weakness of the feature at 1.65\,$\upmu$m (where detection requires SNRs higher than those tested in this paper, i.e. $> 40$). 

From a transit perspective, targeting the \textit{JWST} Mid-Infrared Instrument (MIRI)  wavelength range for our simulations was motivated by \citet{Fauchez_etal_2020}, who highlighted the detectability of an \ce{O2} feature in the mid-infrared (at 6.3\,$\upmu$m) attributed to collision-induced absorption (CIA). The strength of the \ce{O2} detection could be impeded by the nearby water vapor feature at 6.27\,$\upmu$m at a high enough \ce{H2O} abundance, but here we opted to simulate a dry/cold atmospheric terminator scenario for a modern Earth-like TRAPPIST-1e analog consistent with prior studies \citep{Fauchez_etal_2020,Pidhorodetska_2020}. There is also a strong \ce{CH4} feature at 7.66\,$\upmu$m in this wavelength range that would enhance an \ce{O2}-\ce{CH4} chemical disequilibrium detection. 

\subsection{Direct Imaging Simulations}
\label{DIS}
\begin{figure}[H]
    \centering
    \includegraphics[scale=0.80]{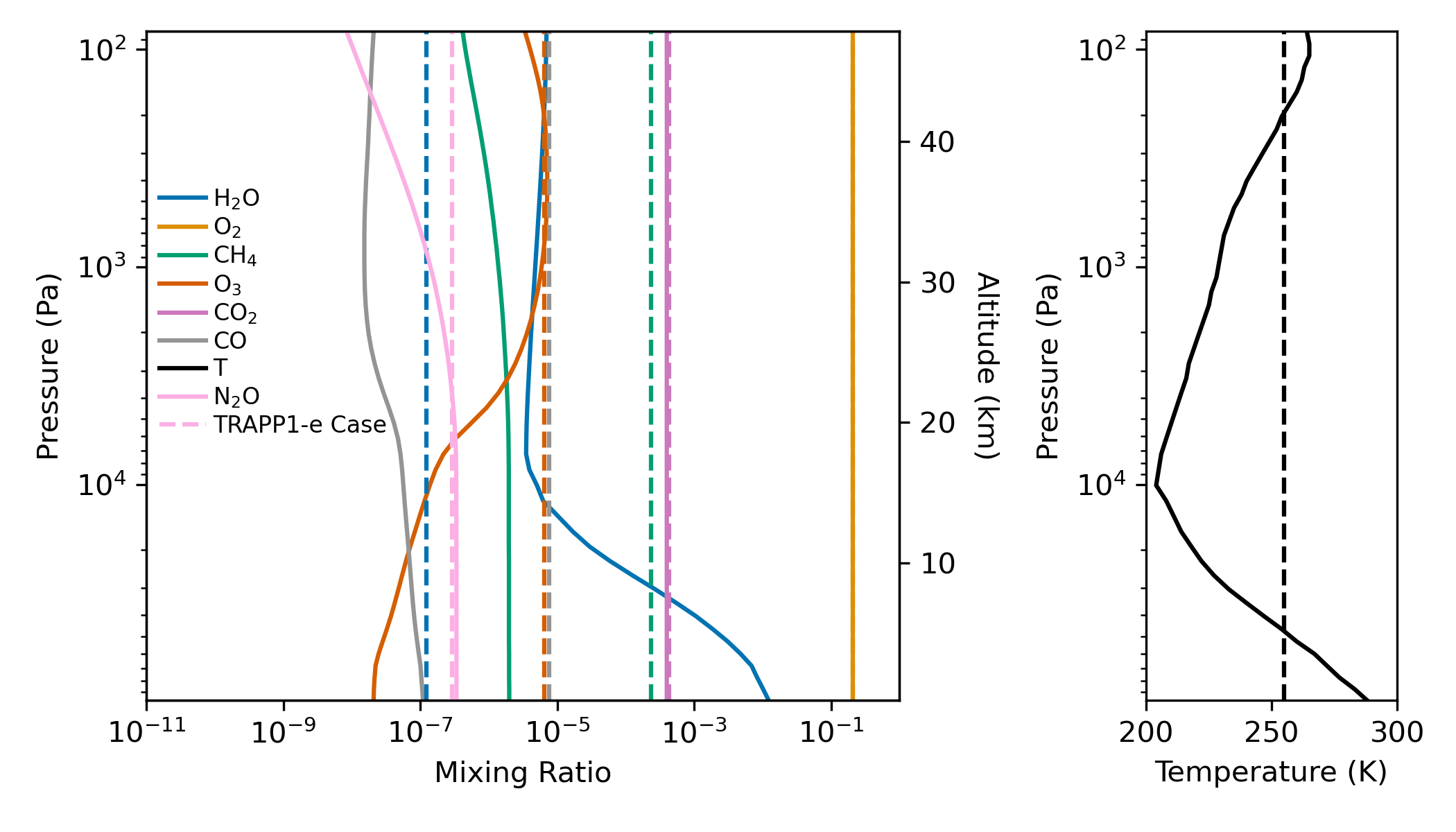}
    \caption{Altitudinal atmospheric profiles for the modern Earth-Sun case (solid lines) and modern Earth-M dwarf case (dashed lines).}
    \label{atm_porfiles}
\end{figure}

In the retrievals, we simulated reflected light observations of an evenly spaced grid of signal-to-noise ratios (SNR) of 10, 20 and 40 with constant noise specified at V-band (0.55\,$\upmu$m) consistent with previous Earth retrieval studies \citep{Feng_2018} and exoplanet decadal studies \citep{LUVOIR_Study_2018,HabEx_Study_2018}. Each retrieval inference assumes an isothermal atmosphere, constant volume mixing ratios for gas phase species, and clouds are modeled a mixture of liquid water and ice. To model the Modern Earth-like atmospheric state, altitudinally dependent profiles were derived from photochemical-climate coupled modeling with \texttt{Atmos}. \texttt{Atmos} can simulate a wide range of planetary atmospheric environments \citep[e.g.,][]{arney_pale_2016,Arney_pale_2017,Teal_2022}  and calculates self-consistent steady state profiles of chemical species using 1-dimensional plane parallel hydrostatic equilibrium calculations. Below, Table \ref{table:Atmos} shows the relevant parameter values and surface mixing ratios of chemical species that were used to generate modern Earth spectra that were then retrieved on. Figure \ref{atm_porfiles} presents the atmospheric profiles of each chemical species alongside with the temperature-pressure profiles for both atmospheric cases. The modern Earth-Sun scenario assumed altitude dependent profiles whereas the modern Earth-M dwarf scenario used constant profiles, as detailed in the following section.

\begin{table}[H] \centering \begin{tabular}{c c c} \hline \hline
Parameter & Description & Surface Value \\ 
\hline

O$_2$ & Molecular oxygen mixing ratio & 0.21\\ 
H$_2$O & Water vapor mixing ratio & $1.23 \times 10^{-2}$\\ 
O$_3$ & Ozone mixing ratio& $2.09 \times 10^{-8}$\\ 
CO$_2$ & Carbon dioxide mixing ratio & $4.00 \times 10^{-4}$\\ 
CH$_4$ & Methane mixing ratio & $2.00 \times 10^{-6}$\\ 
N$_2$O & Nitrous oxide mixing ratio & $3.39 \times 10^{-7}$\\ 
CO & Carbon monoxide mixing ratio & $1.10 \times 10^{-7}$\\ 
$T_0$ (K) & Atmospheric Temperature & 288\\
$P_0$ (Pa) & Surface pressure & $1 \times 10^{5}$\\

\hline 
\end{tabular} \caption{Spectral input atmospheric parameter values calculated by \texttt{Atmos}.} \label{table:Atmos} \end{table}

\begin{figure}[H]
    \centering
    \includegraphics[scale=0.55]{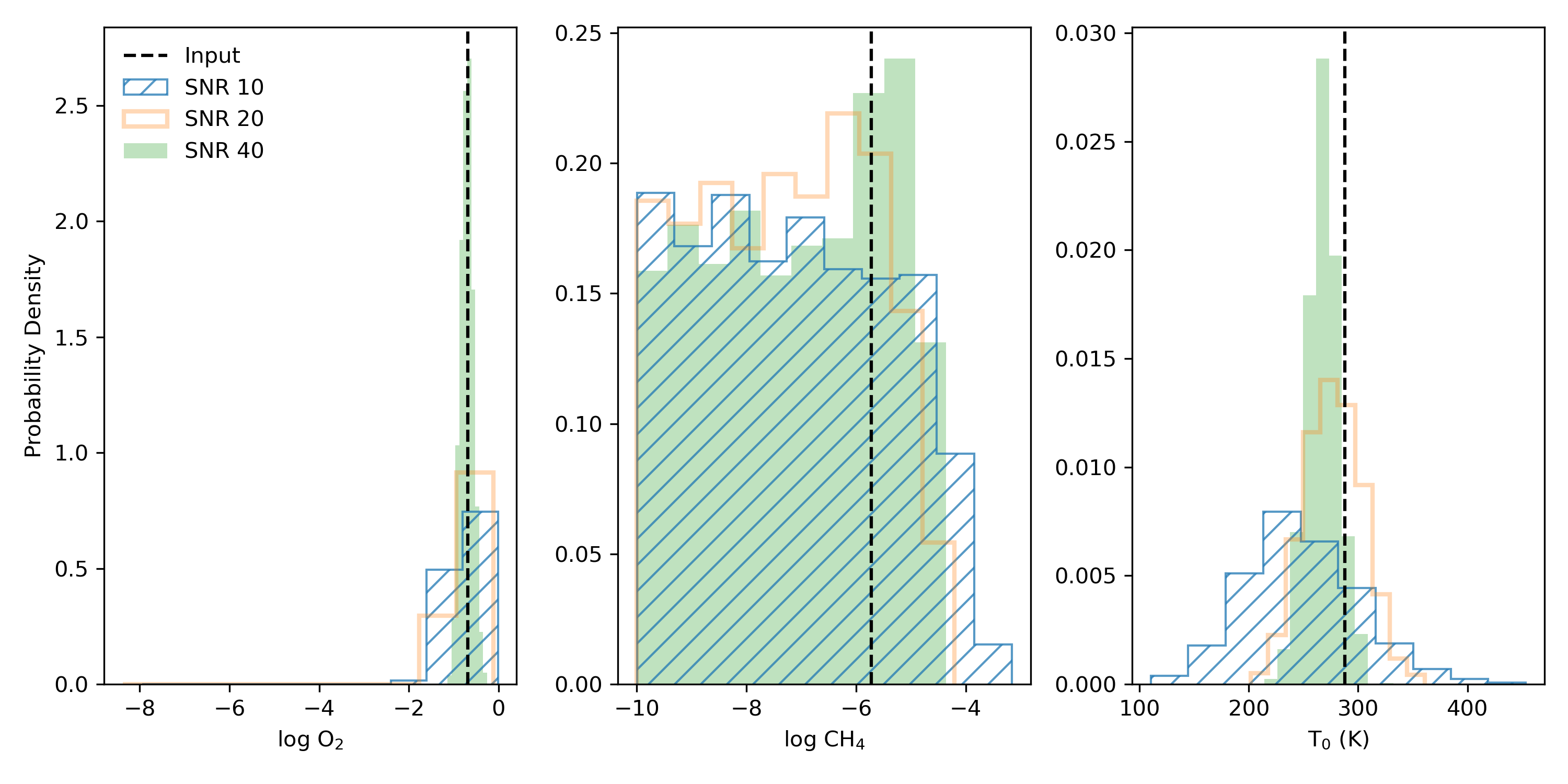}
    \caption{Marginal posterior results for \ce{O2}, \ce{CH4} and $T_0$ derived from simulated direct imaging observations at SNRs of 10 (blue hatched), 20 (orange), and 40 (green). Black vertical dashed lines indicate the input value for each parameter.}
    \label{EP_Refl}
\end{figure}

In Figure \ref{EP_Refl} the marginal posterior distributions for the abundances of \ce{O2}, \ce{CH4} are shown along with atmospheric temperature constraints for a modern Earth like planet imaged in reflected light. The distributions for all three of these parameters were randomly sampled for the available Gibbs free energy calculation as part of coupling the retrievals to the thermodynamics model. The \ce{O2} abundance was able to be constrained at all SNRs tested to within about an order of magnitude. The \ce{CH4} however, is extremely difficult to constrain in reflected light at these modern Earth abundances and in all three observing scenarios we obtain only upper limits. We achieved reasonable atmospheric temperature constraints at all three observational scenarios and these constraints are likely driven by the temperature-dependent shapes of \ce{H2O} features across the visible to near infrared range.

\begin{figure}[H]
    \centering
    \includegraphics[scale=0.55]{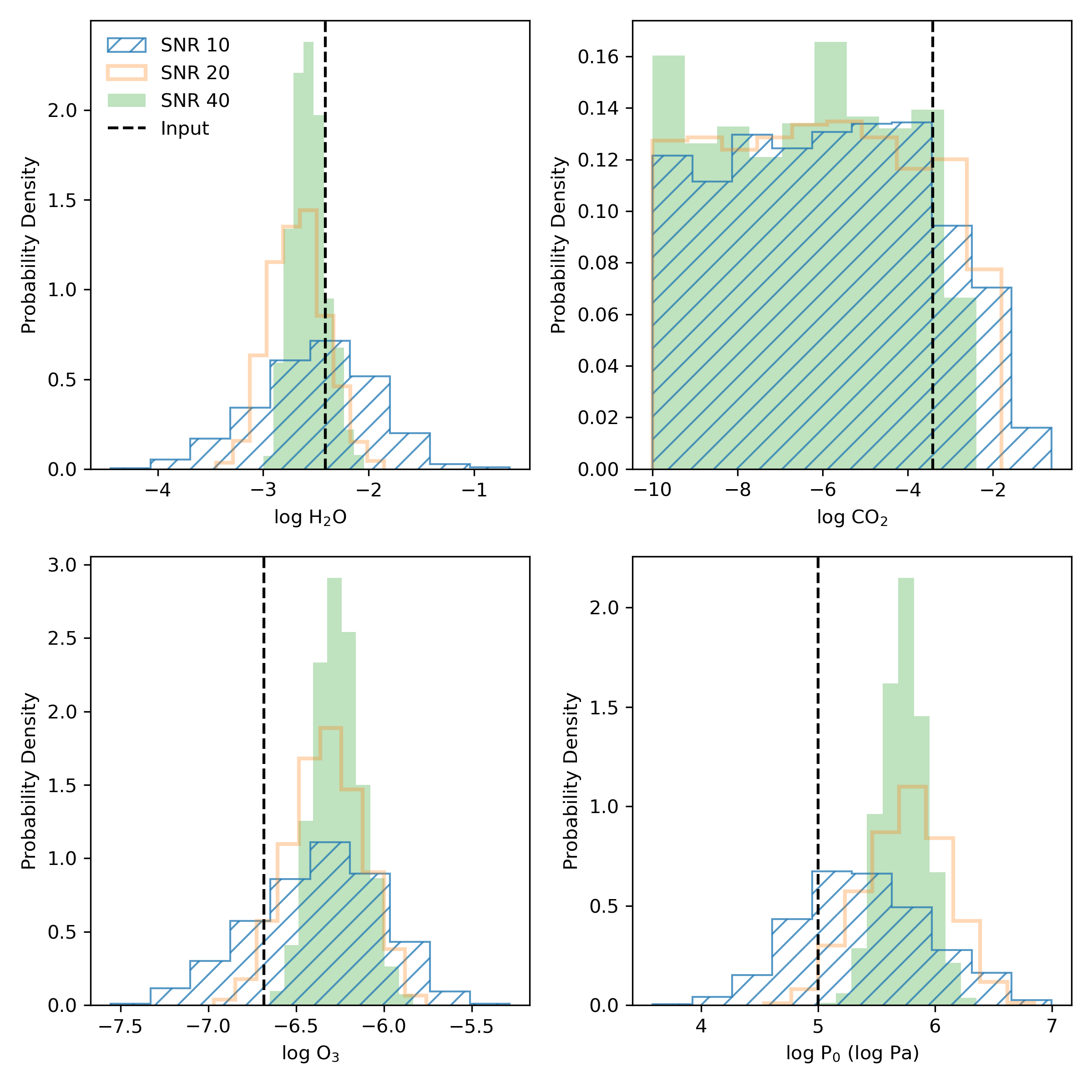}
    \caption{Marginal posterior results for \ce{H2O}, \ce{CO2} \ce{O3}, and P$_0$ derived from simulated direct imaging observations at noise instances of 10 (blue hatched), 20 (orange) and 40 (green) SNR. Input values are indicated with a black vertical dashed line.}
    \label{NEP_Refl}
\end{figure}

The marginal posterior distributions for \ce{H2O}, \ce{CO2}, \ce{O3}, and P$_0$ are shown in Figure \ref{NEP_Refl} at each simulated noise instance (10, 20, and 40 SNR). While each of these parameters are randomly sampled for the chemical disequilibrium calculation, they each have a negligible affect on the overall available Gibbs free energy of the atmosphere. However, these parameters are important to quantify on exoplanets because they provide information about habitability and atmospheric context. We are only able to determine an upper limit constraint on \ce{CO2} for even the highest SNR tested here.  The \ce{O3} abundance retrieved is biased slightly above the input value, which is taken to be the column average. Because the \ce{O3} mixing ratio varies with altitude and the retrieval forward model assumes isoprofiles, this introduces bias in the resultant posteriors that are most evident at high SNRs. The marginal posterior distributions for P$_0$ are also biased high by about an order of magnitude due to the isoprofile and isothermal assumptions made in the retrieval model. Refer to Figures \ref{ME_refl_SNR10}, \ref{ME_refl_SNR20}, and \ref{ME_refl_SNR40} in the Appendix for the complete 14-parameter corner plots corresponding to the SNR 10, 20, and 40 cases, respectively. To verify this, an additional retrieval assuming constant profiles  was conducted for the SNR 40 case, which exhibited the most significant biases. When constant profiles were assumed, biases in these parameters are effectively removed (Figure \ref{ME_self_refl_SNR40} in the Appendix).   

Figure \ref{ME_Refl_Gibbs} shows the available Gibbs free energy in units of Joules per mole of atmosphere for a modern Earth-like planet observed in reflected light. The resulting Gibbs free energy is inferred from simulated reflected light observations performed at SNRs of 10 (blue hatched), 20 (orange), and 40 (green filled). Also plotted for reference are the available Gibbs free energy values for Mars (red dashed line) and the modern Earth atmosphere-ocean system (blue dashed line). The black dashed line represents the truth (or input) value for the chemical disequilibrium in Earth's atmosphere. The peaks in each distribution are biased low by about an order of magnitude due to the extension of the \ce{CH4} marginal distribution down to the prior-imposed lower limit. Obtaining strong available Gibbs free energy constraints for modern Earth in reflected light is very challenging given the difficulty in quantifying \ce{CH4} abundance. Upper limits could be placed on the available Gibbs free energy in these cases since the posteriors minimally overlap with the referenced Mars, and Earth atmosphere-ocean values.

\begin{figure}[H]
    \centering
    \includegraphics[scale=0.7]{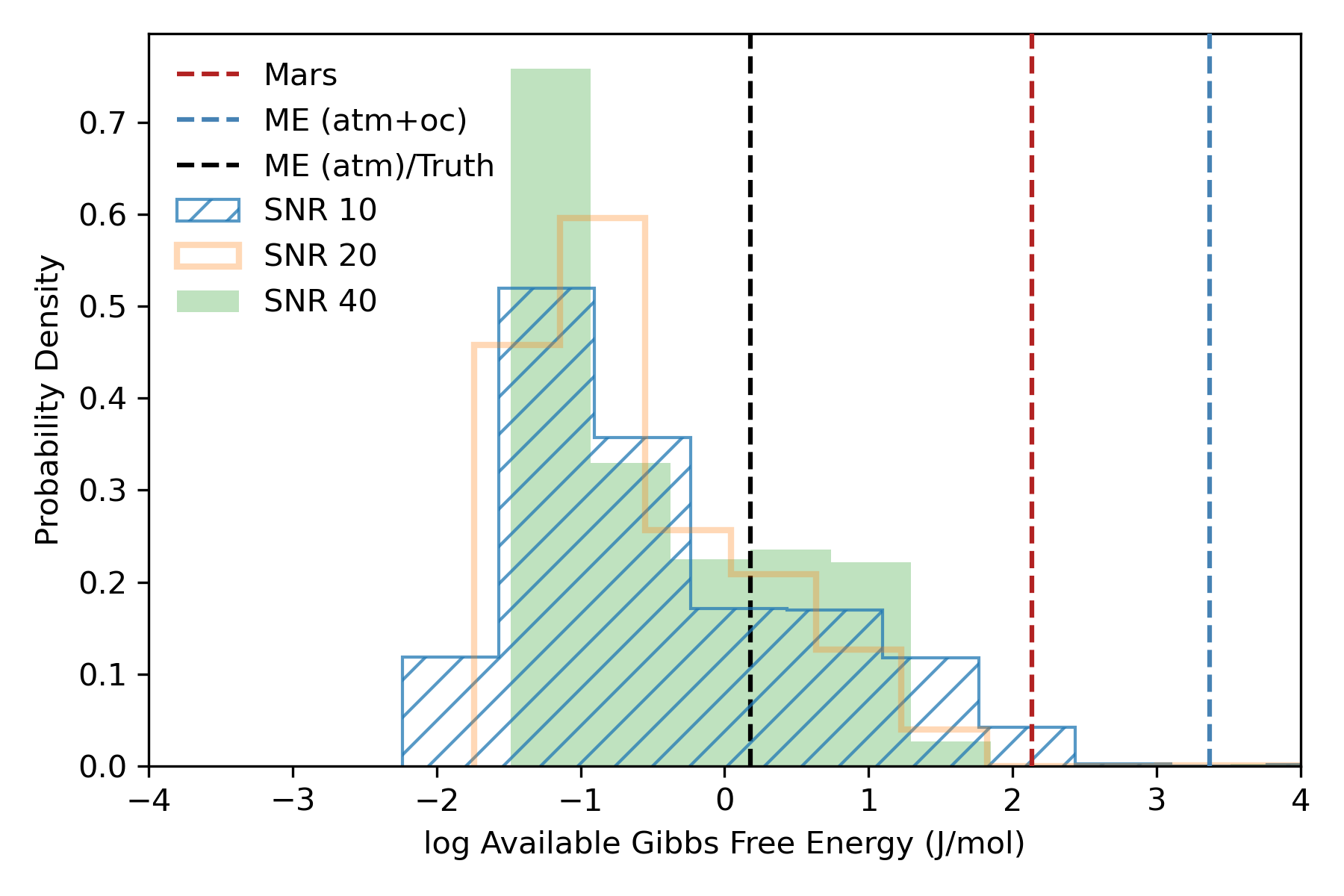}
    \caption{Available Gibbs free energy posterior derived from simulated reflected light observations of a modern Earth-like exoplanet observed in reflected light with simulated observations performed at SNRs of 10 (blue hatched), 20 (orange) and 40 (green filled). Vertical dashed lines indicate the available Gibbs free energy values for Mars (red), modern Earth's atmosphere-ocean system (blue), and the true value (black).}
    \label{ME_Refl_Gibbs}
\end{figure}

\subsection{JWST/MIRI Transit Simulations}
\label{MIRI_T_S}
The following results outline available Gibbs free energy inferences for a planet with modern Earth-like gas fluxes orbiting a late-type M dwarf. The simulated transit observations were modeled after the \textit{JWST} \textit{MIRI} instrument and spanned a wavelength range from 5 to 12\,$\upmu$m at a fixed resolving power of 40. Although \textit{JWST's}  \textit{NIRSpec} instrument could infer a \ce{CO2},\--\,\ce{CH4} driven disequilibrium, the challenge of constraining \ce{O2} at modern Earth abundances \citep{Fauchez_etal_2020,Krissansen_Totton_TRAPPIST_2018,Lustig_Yeager_2019,Wunderlich_2019} motivated the decision to focus on simulating \textit{MIRI} observations. We modeled a modern Earth-like TRAPPIST-1e analog with a planetary radius of 0.9\,R$_\oplus$, a planetary mass of 0.7\,M$_\oplus$, and an orbital distance of 0.02 AU. The stellar properties were modeled after TRAPPIST-1, a late type M dwarf with a stellar radius of 0.117\,R$_\odot$ and an effective temperature of 2560\,K. In contrast to the reflected light retrievals, we opted to explore cloud free atmospheric inferences in this preliminary study. The composition of the atmosphere assumed constant gas mixing ratio profiles taken to be consistent with 3D model predictions for the cold terminator of an Earth-like TRAPPIST-1e from \cite{Pidhorodetska_2020}. The values for each species are outlined in Table \ref{table:species}. The mixing ratio values are taken to be the logarithmic geometric mean of the stratospheric values from 30 km and 60 km. For \ce{CH4} and \ce{N2O} in particular, their assumed mixing ratios are the predicted values associated with biological production. For pressure and planetary radius inferences the reference radius was set to 1\,bar (10$^5$\,Pa) for all the simulated transit observations.

\begin{table}[h!] \centering \begin{tabular}{l c c} \hline \hline Species & Mixing Ratio & Citation \\ 
\hline 
O$_2$ & $2.09 \times 10^{-1}$ & \cite{Pidhorodetska_2020}\\ 
H$_2$O & $1.24 \times 10^{-7}$ & \cite{Pidhorodetska_2020}\\ 
O$_3$ & $6.51 \times 10^{-6}$ & \cite{Pidhorodetska_2020}\\ 
CO$_2$ & $4.30 \times 10^{-4}$ & \cite{Pidhorodetska_2020}\\ 
CH$_4$ & $2.36 \times 10^{-4}$ & Calculated from Atmos\\ 
N$_2$O & $2.90 \times 10^{-7}$ & Calculated from Atmos\\ 
CO & $7.63 \times 10^{-6}$ & \cite{Pidhorodetska_2020}\\ 
\hline 
\hline
\end{tabular} \caption{Mixing ratios and citations for various chemical species. Constant profiles were used for each species to generate the corresponding transit spectrum.} \label{table:species} \end{table}

\begin{figure}[H]
    \centering
    \includegraphics[scale=0.55]{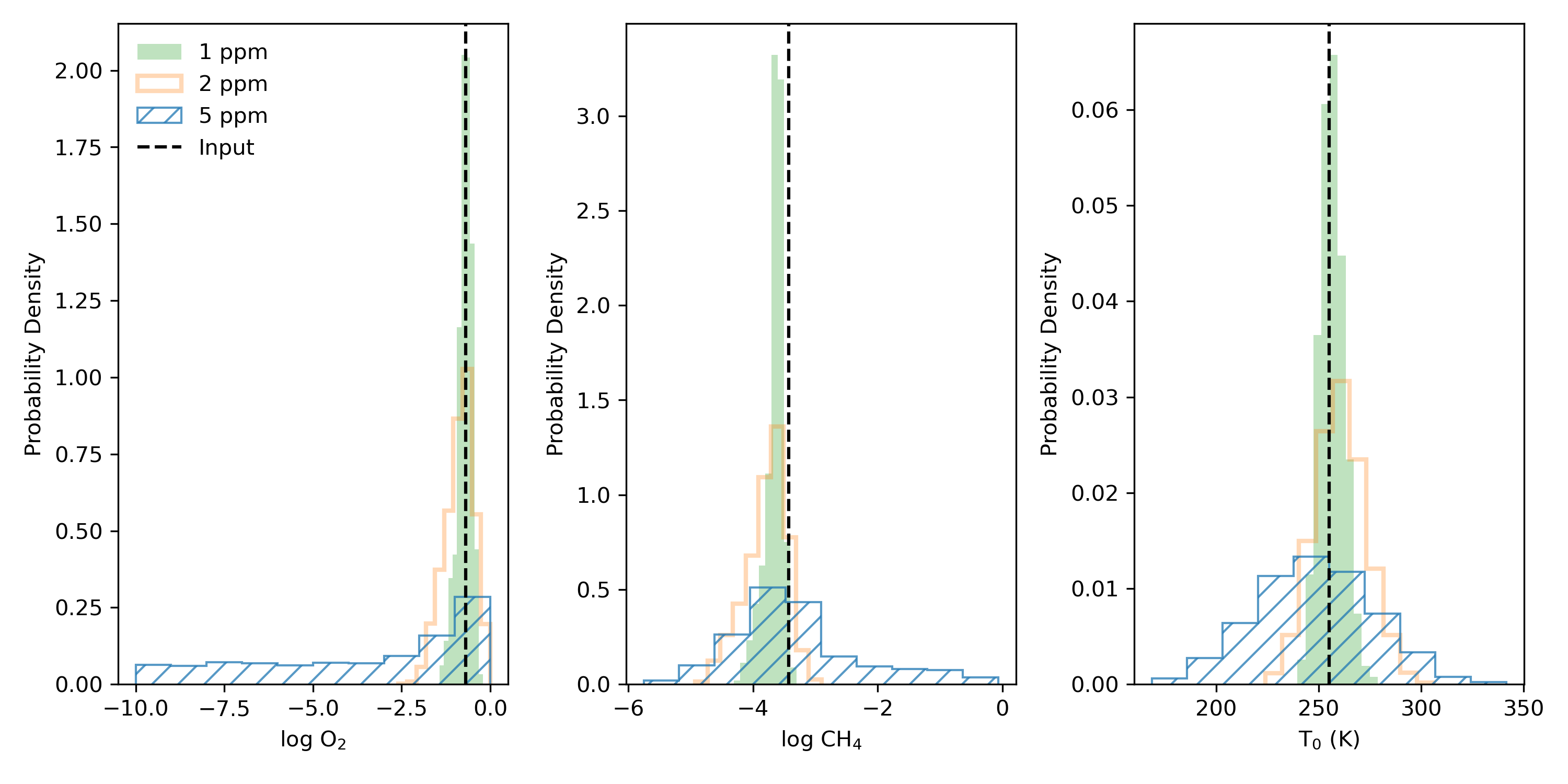}
    \caption{Marginal posterior results for \ce{O2}, \ce{CH4} and $T_0$ derived from simulated transit observations at 1 (green filled) 2 (orange) and 5 (blue hatched) ppm noise levels. Input values are indicated with a black vertical dashed line.}
    \label{EP_MIRI}
\end{figure}

In Figure \ref{EP_MIRI} the marginal posterior distributions are shown for the abundances of \ce{O2} and \ce{CH4} as well as for atmospheric temperature derived from transit observations at 1 (green filled), 2 (orange) and 5 (blue hatched) ppm noise. Each of these parameters substantially contribute to the available Gibbs free energy similarly to the previous Earth-Sun case. While the \ce{O2} input abundance remains consistent with the Earth-Sun scenario, more atmospheric \ce{CH4} is allowed to accumulate in this Earth-M dwarf scenario because the incident UV spectrum and resultant atmospheric photochemistry are different in comparison to an Earth twin \citep[e.g.,][]{Segura_2005,Rugheimer_2015,Arney_2019}. At 1\,ppm and 2\,ppm, the full extent of the marginal posterior distributions for each parameter are constrained to within an order of magnitude. At 5\,ppm the distributions for each parameter broaden markedly so that, most significantly, \ce{O2} goes undetected. 

\begin{figure}[H]
    \centering
    \includegraphics[scale=0.55]{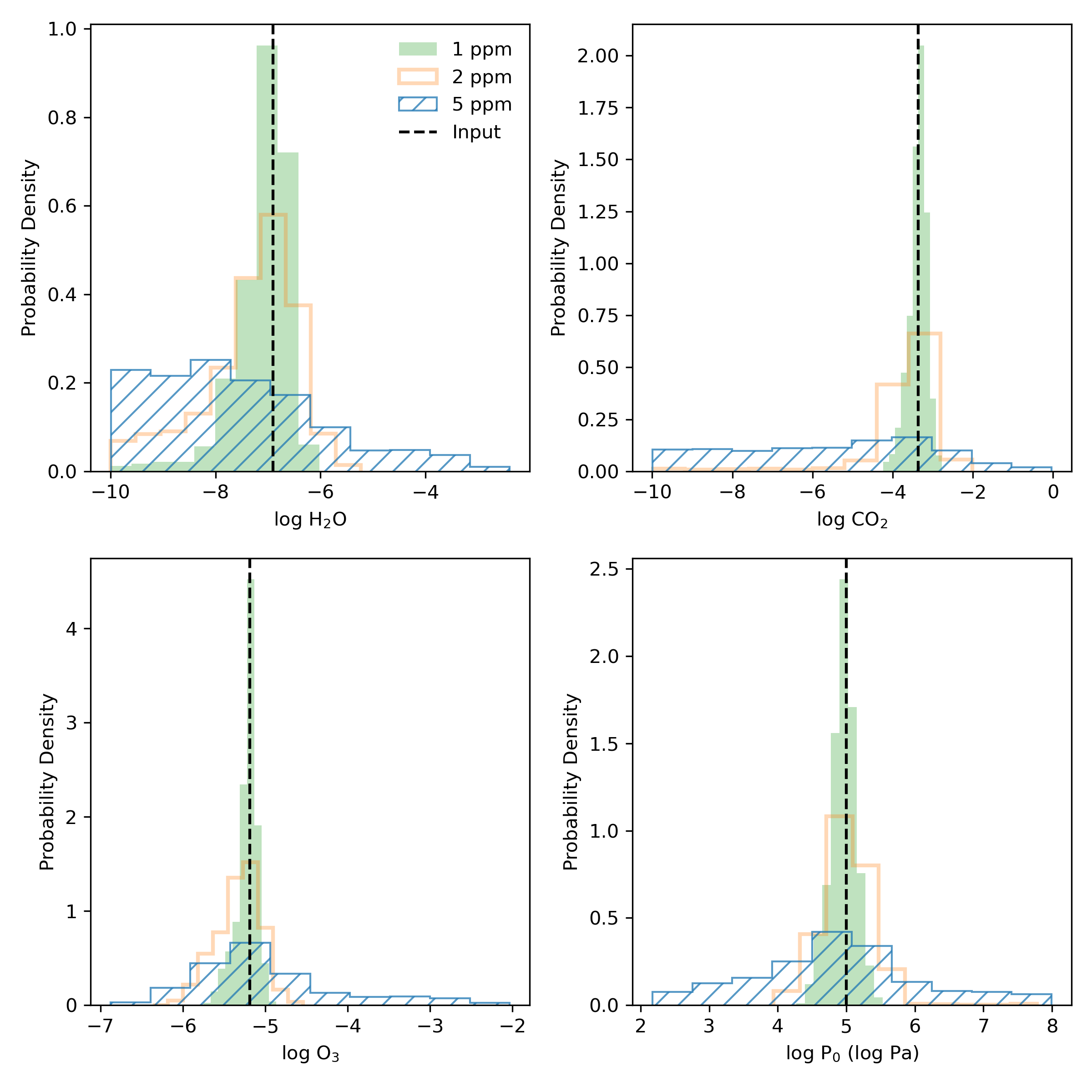}
    \caption{Marginal posterior results for \ce{H2O}, \ce{CO2}, \ce{O3}, and $P_0$ derived from simulated transit observations performed at noise instances of 1 (green filled), 2 (orange), and 5 (blue hatched) ppm. Input values are indicated with a black vertical dashed line.}
    \label{NEP_MIRI}
\end{figure}

Figure \ref{NEP_MIRI} shows the posterior distributions for the retrieved abundances of \ce{H2O}, \ce{CO2}, \ce{O3}, and the surface pressure for MIRI transit observations at the studied noise levels. At 5\,ppm it is again difficult to constrain these parameters. For \ce{H2O}, it becomes challenging to put a lower limit on the abundance because its primary absorption feature, like the \ce{O2} CIA feature at these wavelengths, is weaker than the noise level. And particularly for surface pressure, which is defined by an effective radius set at $10^5$\,Pa, there is sensitivity to the surface at noise levels $<$ 5\,ppm, but at 5\,ppm noise the surface pressure is no longer constrained and deep atmospheric solutions can no longer be ruled out. Figures \ref{ME_MIRI_5ppm}, \ref{ME_MIRI_2ppm}, and \ref{ME_MIRI_1ppm}, in the Appendix show the full 9-parameter corner plots for the corresponding 5\,ppm, 2\,ppm, and 1\,ppm cases respectively.

\begin{figure}[H]
    \centering
    \includegraphics[scale=0.7]{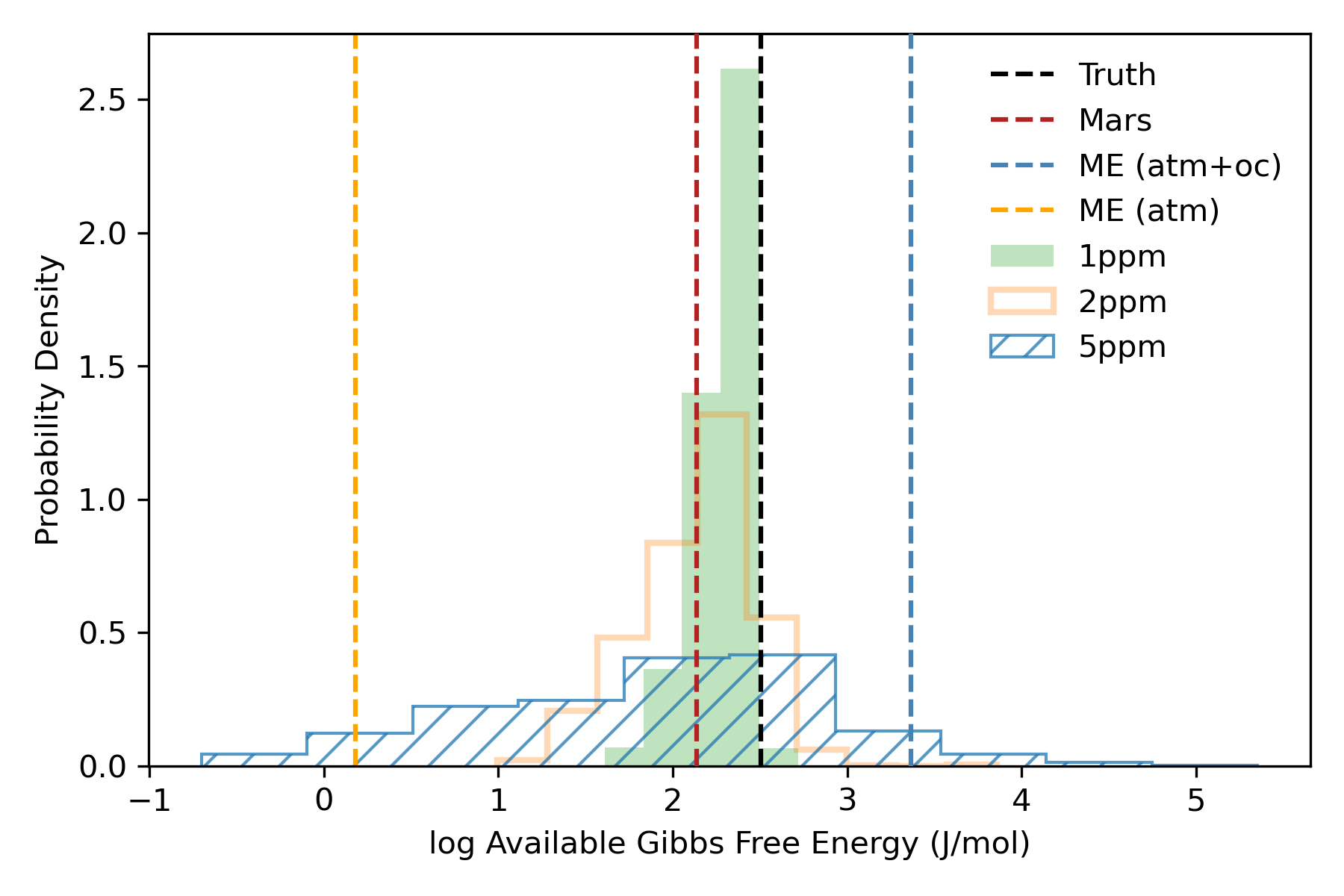}
    \caption{Available Gibbs free energy posterior derived from simulated transit observations of a modern Earth-like exoplanet with \textit{JWST} MIRI instrumentation. For comparison, the available Gibbs free energies of worlds around the sun are provided. Shown are the available Gibbs free energies for Mars (red vertical dashed line), the modern Earth atmosphere-ocean system (blue vertical dashed line), and the modern Earth atmospheric system (vertical orange line). The vertical black dashed line represents the predicted free of a modern Earth M dwarf analog and is taken to be the ``truth" value for this planetary scenario.}
    \label{ME_MIRI_Gibbs}
\end{figure}

Figure \ref{ME_MIRI_Gibbs} shows the inferred available Gibbs free energy for the atmosphere of a modern Earth orbiting an M dwarf inferred from simulated transit observations with MIRI. Each observation was simulated at noise levels of 1 (green filled), 2 (orange) and 5 (blue hatched) ppm. For comparison, the available Gibbs free energies of solar system bodies (i.e., Mars, and Earth's atmosphere/atmosphere-ocean system) are plotted as well for comparison as in the reflected light study. The predicted free energy in this case is higher in comparison to the Earth-Sun scenario given the higher amounts of \ce{CH4} that accumulate in this case. The 1\,ppm observation provided the most optimistic constraints on each of the observed parameters, which ultimately leads to tight constraints on the inferred available Gibbs free energy. At the 2\,ppm noise level, good constraints on the available Gibbs free energy are obtained where the distribution is peaked near the true value. At 5\,ppm the available Gibbs free energy goes unconstrained and implies the Gibbs free energy cannot be constrained at this level of noise.

\section{Discussion}\label{sec:disc}
Chemical disequilibrium is an important metric to include in the search for life alongside biosignature gas detection strategies because it could be a more agnostic sign of life, and we demonstrate the process to remotely inferring chemical disequilibrium signals from information derived from spectroscopic retrievals. The \texttt{rfast} retrievals were coupled to the thermodynamics model by randomly sampling the retrieval parameter posteriors and passing those randomized inputs to the thermodynamics model in order to compute a marginal posterior distribution for the available Gibbs free energy, which is the metric for quantifying chemical disequilibrium. By coupling the \texttt{rfast} retrievals to the thermodynamics model, the marginal posterior distribution for the available Gibbs free energy can shed light on how observational uncertainty can influence the resulting available Gibbs free energy distributions. Notably, for an Earth-like exoplanet atmosphere, chemical disequilibrium is maintained through the co-existence of \ce{O2} and \ce{CH4} and is highly sensitive to the observational uncertainty of these gases and atmospheric temperature. We find that in reflected light, the available Gibbs free energy is challenging to detect for observing scenarios at SNRs of 10, 20, and 40 and is mainly limited by the uncertainty of \ce{CH4}, which is difficult to constrain at modern Earth abundances. In fact, the minimum SNR required to achieve a \ce{CH4} detection at a modern Earth abundance is an exorbitantly high value of 192 for the NIR feature spanning 1.64 - 1.7 $\mu$m. This was computed using a spectral differencing approach and summing the wavelength dependent SNR over the extent of the feature to determine the minimum SNR needed for robust constraints. For the case of a modern Earth-like TRAPPIST-1e analog orbiting an M dwarf, we see improved constraints on the available Gibbs free energy albeit at extremely low noise levels and assuming a simplified cloud free scenario.

Retrieved parameters like pressure and the abundances of \ce{H2O}, \ce{CO2}, and \ce{O3} do not contribute substantially to the overall available Gibbs free energy. However, the inclusion of \ce{H2O}, \ce{CO2}, and \ce{O3} are crucial for maintaining the mass balance in the Gibbs free energy optimization and including all of these parameters in the atmospheric retrievals is important for generally characterizing the atmospheric state of the planet. For example, planetary surface pressure, \ce{H2O}, and \ce{CO2} help to inform our understanding of climate and rule out certain atmospheric regimes in the retrieval parameter space. Additionally, \ce{O3} is a photochemical byproduct of \ce{O2} and in practice can be used to infer the presence of oxygen in a planetary atmosphere since \ce{O3} can remain detectable at UV wavelengths even at low \ce{O2} abundances; a possibility we do not explicitly include in our retrieval because in principle \ce{O3} could be used to better constrain \ce{O2}, and thereby the available Gibbs free energy \citep{Meadows_2017,Meadows_2018,Schwieterman_2018,kozakis_2022}. We removed a number of chemical species that were included in the default thermodynamic calculation (i.e., \ce{Ar}, \ce{Ne}, \ce{He}, \ce{Kr}, \ce{Xe}, \ce{H2}, \ce{N2O}, \ce{CO}, and \ce{HCl}) from the retrieval analysis. These species did not have a substantial influence on the available Gibbs free energy; either because they were not in chemical disequilibrium (which was the case for the noble gases) or their small relative abundances had a negligible impact. Our calculated available Gibbs free energy for the modern Earth-sun scenario is $\sim$\,1\,J\,mol$^{-1}$ and this is in line with previous studies \citep{krissansen-totton_detecting_2016}.   

In general, we found that constraining the chemical disequilibrium for modern Earth in reflected light around a G dwarf, or in transit around TRAPPIST-1, is challenging. In reflected light, available Gibbs free energy inferences of a modern Earth twin are hindered by poor \ce{CH4} abundance constraints for simulated observations at $<$\,SNR 40. While achieving a \ce{CH4} detection at the modern Earth abundance would take an exorbitant amount of integration time to reach the minimum SNR requirement of 192, Gibbs free energy inferences may be more achievable for modern Earth-like planets orbiting M-type or K-type stars due to the extended photochemical lifetime of \ce{CH4}. At greater SNRs, performing retrieval analyses, which generally adopt isothermal and constant mixing ratio profiles, becomes a poor assumption for a realistic modern Earth scenario in which these atmospheric profiles vary with altitude. For the SNR 40 observation, the pressure for example, becomes biased toward higher values by a factor of 5 alluding to systematic effects resulting from the retrieval assumptions. This was verified with the retrieval simulation that assumed constant profiles, in which case the biases on the aforementioned parameters were effectively removed.

Inferring the available Gibbs free energy of a modern Earth TRAPPIST-1e analog scenario was potentially more promising given the heightened \ce{CH4} abundances in this planetary context. In comparison to the modern Earth-Sun available Gibbs free energy (which is $\sim$\,1\,J\,mol$^{-1}$), the available Gibbs free energy of the modern Earth-M dwarf case was substantially higher at $\sim$\,320\,J\,mol$^{-1}$. To determine the requirements necessary to make a strong chemical disequilibrium inference, clouds were omitted and a reference radius was set at 1\,bar to maximize our ability to remotely sense the deep atmosphere, which is especially important for retrieving chemical species abundances. The presence of high altitude clouds, or reduced sensitivity to species abundances that vary with altitude could introduce sources of bias in the resulting available Gibbs free energy posteriors. Here, modeling a simple scenario with constant stratospheric mixing ratio profiles and omitting clouds mitigates bias for the pressure constraints, and abundance constraints for species like \ce{CH4} and \ce{O2}. Additionally, the resulting constraints on the available Gibbs free energy are tight with less than an order of magnitude spread on the 1\,ppm noise case. However, this simplified scenario may not be realistic especially with regard to the absence of clouds in an atmospheric regime with \ce{H2O} present. Furthermore, overcoming an observational noise floor to achieve observations at several ppm is likely implausible for \textit{JWST} \citep{Greene_2016,Rustamkulov_2022}.

To address potential false-positive scenarios for generating chemical disequilibrium signals abiotically, we compared the available Gibbs free energies of our Earth-like cases to Mars' repoted value from \citet{krissansen-totton_detecting_2016}. Notably, Mars represents a unique case within the solar system due to its larger atmospheric Gibbs free energy relative to Earth's where Mars' chemical disequilibrium is primarily generated through abiotic photolysis of \ce{CO2}. Distinguishing between biotic and abiotic cases is critical. The results presented here suggest that establishing robust upper limit constraints on available Gibbs free energy could further differentiate these scenarios, particularly in the modern Earth-Sun context. For the modern Earth-like TRAPPIST-1e case, distinguishing it from the Mars case is more challenging, as the higher available Gibbs free energy in the M dwarf scenario is comparable in magnitude to that of Mars. Critically, identifying the specific chemical species driving the disequilibrium signal would provide essential contextual information for distinguishing abiotic from biotic signals.

There are additional mechanisms for contributing to chemical disequilibrium overall including the contributive Gibbs free energy from atmosphere-ocean interactions which drive a multiphase chemical disequilibrium between \ce{N2}, \ce{O2} and liquid water for modern Earth. Thermodynamic systems considering multiphase interactions are excluded in this study and only the chemical disequilibrium driven by atmospheric species is considered. This is because remote observations sensitive enough to discern the oceanic state and its dissolved species is likely not feasible. In practice remotely constrained available Gibbs free energies for Earth-like worlds would likely provide conservative estimates on the magnitude of disequilibrium that could be present and driven by the gaseous abundance inferences made with remote observations. Despite the challenges, the results from this work have critical implications for future exoplanet characterization efforts and the search for life elsewhere. Currently, there are numerous life detection strategies and metrics that are being developed in preparation for future exoplanet characterization missions. With the development of these metrics, it is highly likely that no one metric will provide a robust life detection for every planetary context. A confident life detection could instead come from building a hierarchy of evidence that, taken together, suggests a given planet hosts life. Chemical disequilibrium is one metric we can use in addition to other measures including the search for particular biosignature gases. 

Additionally, exploring the potential to infer the available Gibbs free energy of Earth-like exoplanets provides a valuable opportunity to benchmark retrieval results against ``ground truth" conditions akin to Earth's well-characterized atmosphere. Furthermore, these inference techniques can be integrated in future life detection frameworks to reinforce the reliability of life detections. For instance, an exoplanet exhibiting Earth-like atmospheric abundances of species like \ce{O2}, \ce{O3}, \ce{CH4} etc., might still differ in terms of their thermodynamic state with regard to other parameters such as atmospheric temperature and pressure. In such cases, Gibbs free energy calculations could efficiently incorporate these variables to assess the extent of chemical disequilibrium. Similarly, analyzing a world with intriguing chemistry could illuminate the potential chemical pathways and sources driving its disequilibrium, providing deeper scientific insights. And while inferring the available Gibbs free energy of modern Earth may be challenging in practice, we have an established modeling infrastructure to couple retrievals to a thermodynamics model in order to enable future chemical disequilibrium inferences.

\section{Conclusions}
\label{Conc}
To summarize, our key findings include:
\begin{itemize}
    \item Inferring chemical disequilibrium signatures can be accomplished by loosely coupling a thermodynamics model on the backend of atmospheric retrievals by randomly sampling marginal posterior distributions of relevant parameters. This is important for determining the remote detectability of such biosignatures and their sensitivity to observational uncertainty.
    \item The detectability of modern Earth-like chemical disequilibrium biosignatures is heavily dependent on the observational uncertainty of \ce{O2}, \ce{CH4} and atmospheric temperature.
    \item In the context of direct imaging observations, the ability to infer \ce{O2}-\ce{CH4} disequilibrium is made difficult by \ce{CH4}, which is challenging to constrain for modern Earth-like abundances with simulated observations at SNRs  $<$ 40.
    \item While the chemical disequilibrium produced by an Earth-like planet around an Mdwarf is higher than the Earth-sun case, assuming a simplified scenario, obtaining constraints on the chemical disequilibrium signal would require an exceedingly low observational noise floor. 
\end{itemize}

Although constraining the available Gibbs free energy for modern-Earth like scenarios presents significant challenges, this work demonstrates critical steps toward integrating thermodynamics calculations into retrieval inferences. As a versatile and agnostic metric, Gibbs free energy offers a promising pathway for future exoplanet characterization efforts, bridging information from spectral observations with thermodynamics calculations. By incorporating chemical disequilibrium into the search for life, we can refine biosignature detection strategies and shape the observational priorities of next-generation exoplanet missions, ensuring readiness for both Earth-like and unexpected chemistries revealed by future exoplanet observations.

\acknowledgments
AVY, TDR, JKT, ES, and GA all acknowledge support from the NASA Exobiology Program, Grant No.~80NSSC18K0349. AVY, TDR, EWS, and GA also acknowledge support from NASA's Nexus for Exoplanet System Science Virtual Planetary Laboratory (No.~80NSSC18K0829). AVY would like to acknowledge support from the GSFC Sellers Exoplanet Environments Collaboration (SEEC), which is supported by NASA's Planetary Science Division's Research Program. AVY also acknowledges support from the NASA Fellowship Activity Program awarded through Grant No.~80NSSC21K2064 and support from the NASA Pathways Intern Program administered by the local Pathways Office at NASA Goddard. TDR acknowledges support from the Cottrell Scholar Program administered by the Research Corporation for Science Advancement. ES acknowledges additional support from the NASA Interdisciplinary Consortia for Astrobiology Research (ICAR) Program issued under grant Nos.~80NSSC21K0594 and ~80NSSC21K0905. Some of the Computational analyses were run on Northern Arizona University’s Monsoon computing cluster, funded by Arizona’s Technology and Research Initiative Fund. AVY would like to sincerely thank Dr. Christian Tai Udovicic for their invaluable guidance during the development of our modeling framework in Python and helping to craft solutions that optimized the model's architecture and functionality. Finally, the authors would like to sincerely thank the anonymous reviewer for their insightful comments and feedback during the review process.

%



\software{\texttt{rfast} \citep{Robinson_2023}, \texttt{emcee} \citep{Foreman_Mackey_2013}, Pairwise from MC3 \citep{Cubillos_2017} \texttt{Atmos}: \url{https://github.com/VirtualPlanetaryLaboratory/atmos}, Matlab thermodynamics model: \url{http://www.krisstott.com/code.html}, Python thermodynamics model: \url{https://github.com/Bellatrix12/Python_Equilibrium_Code.git}. Accessible figure data: \url{https://doi.org/10.5281/zenodo.15485517}}




\appendix

\section{Modern Earth-Sun Twin Case}

\begin{figure}[H]
    \centering
    \includegraphics[scale=0.45]{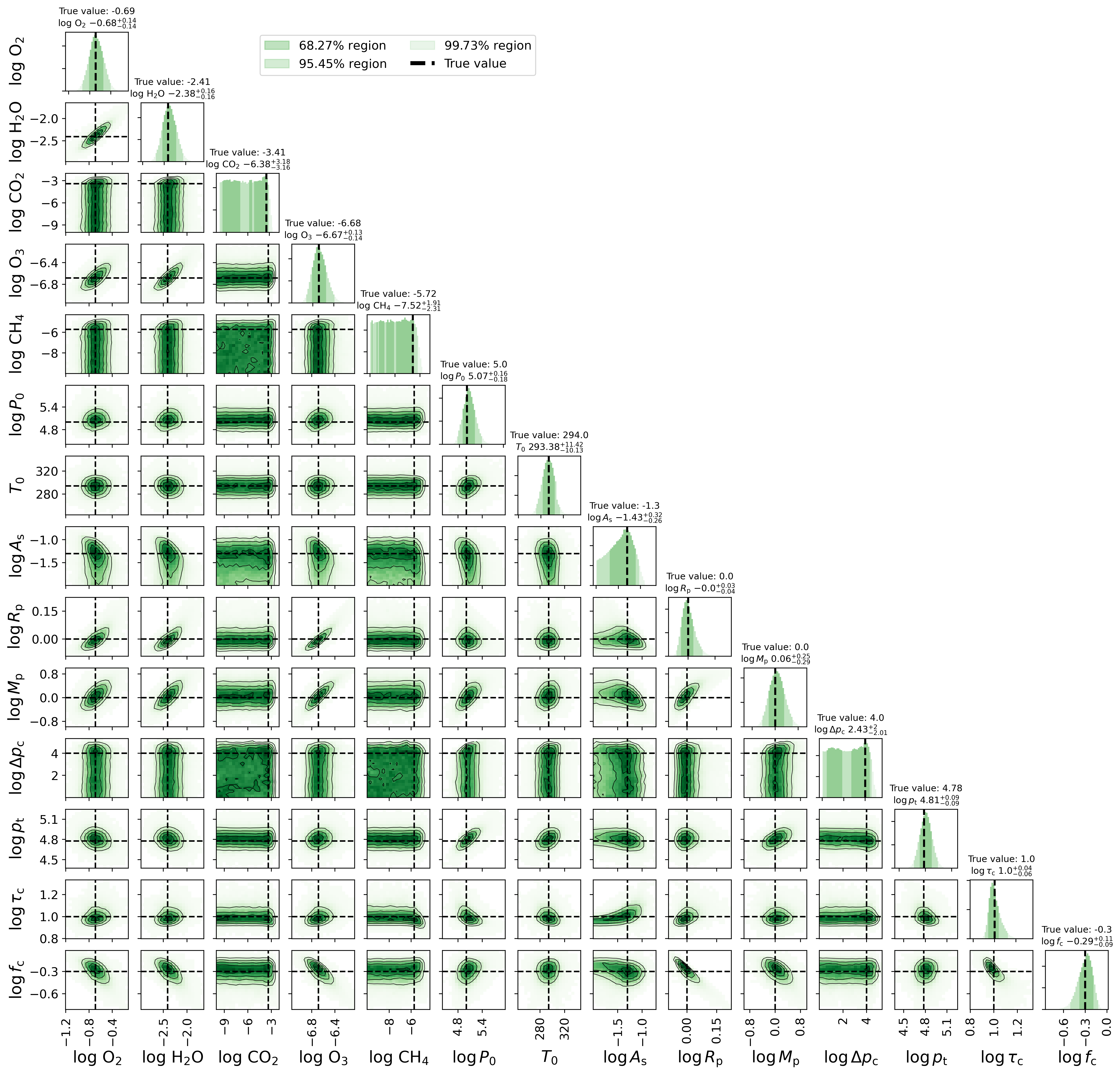}
    \caption{The full 14-parameter corner plot for an SNR 40 simulated observational scenario assuming isoprofiles. One-dimensional marginal posterior distributions are shown along the diagonal with associated truth values and the 16$^{\rm th}$, 50$^{\rm th}$, and 84$^{\rm th}$ percentile values indicated above each marginal distribution. The truth values for each parameter are also indicated by black dashed lines.}
    \label{ME_self_refl_SNR40}
\end{figure}

\begin{figure}[H]
    \centering
    \includegraphics[scale=0.45]{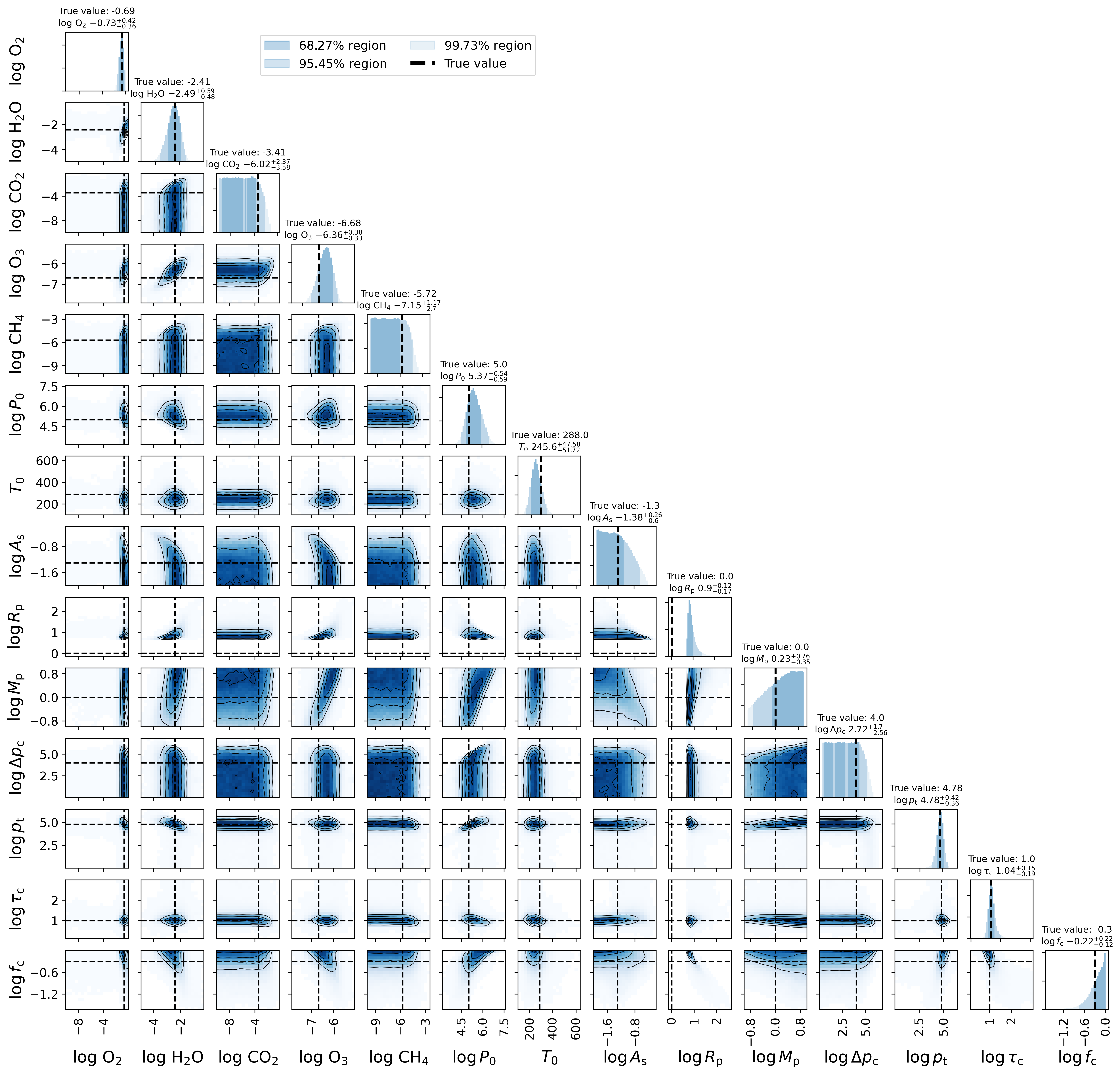}
    \caption{The full 14-parameter corner plot for the SNR 10 simulated observational scenario presented in Section \ref{DIS}. One-dimensional marginal posterior distributions are shown along the diagonal with associated truth values and the 16$^{\rm th}$, 50$^{\rm th}$, and 84$^{\rm th}$ percentile values indicated above each marginal distribution. The truth values for each parameter are also indicated by black dashed lines.}
    \label{ME_refl_SNR10}
\end{figure}

\begin{figure}[H]
    \centering
    \includegraphics[scale=0.45]{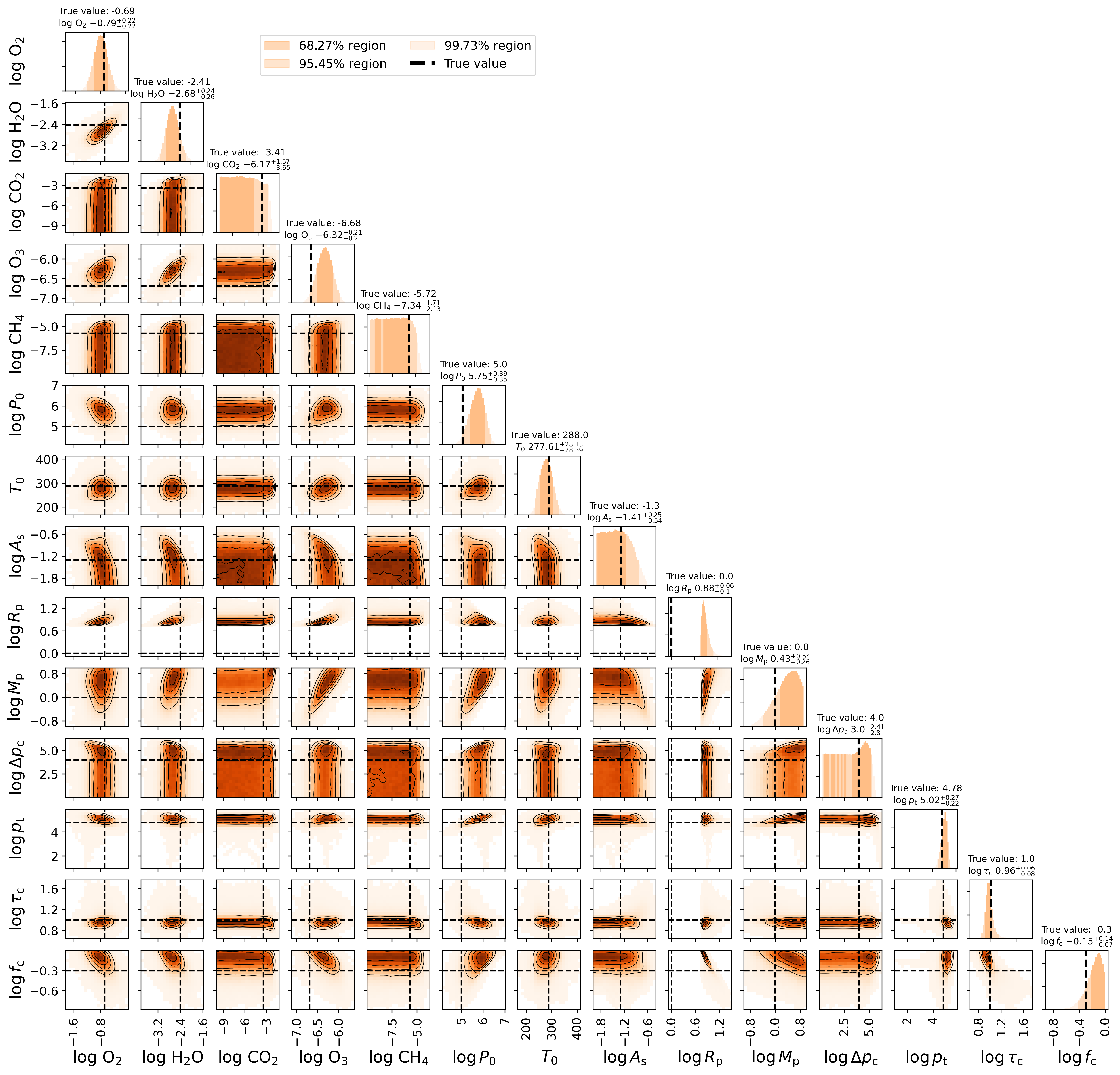}
    \caption{Same as Figure \ref{ME_refl_SNR10} but for SNR 20.}
    \label{ME_refl_SNR20}
\end{figure}

\begin{figure}[H]
    \centering
    \includegraphics[scale=0.45]{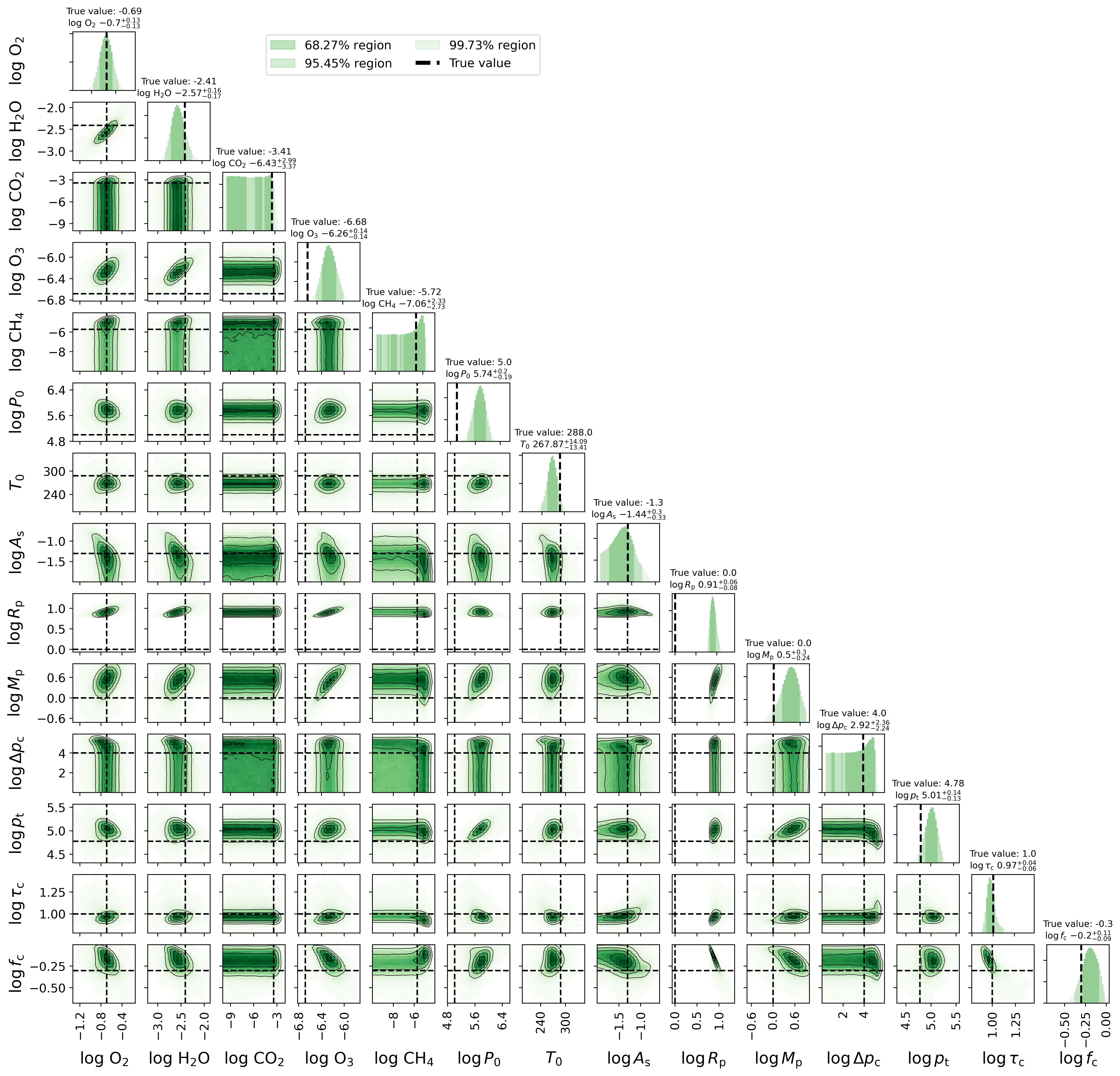}
    \caption{Same as Figure \ref{ME_refl_SNR10} but for SNR 40.}
    \label{ME_refl_SNR40}
\end{figure}

Figure \ref{ME_self_refl_SNR40} depicts a retrieval simulation performed at an SNR of 40 with assumed isoprofiles for gas phase species abundances, temperature, and pressure. Figure \ref{ME_refl_SNR10}, Figure \ref{ME_refl_SNR20}, and Figure \ref{ME_refl_SNR40} depict the full retrieval results for the SNR 10, 20, and 40 observational cases described in Section \ref{DIS}. Above each 1-D posterior, the truth values, and 1-$\sigma$ confidence regions are reported.

\newpage

\section{Modern Earth TRAPPIST-1e Case}
\begin{figure}[H]
    \centering
    \includegraphics[scale=0.45]{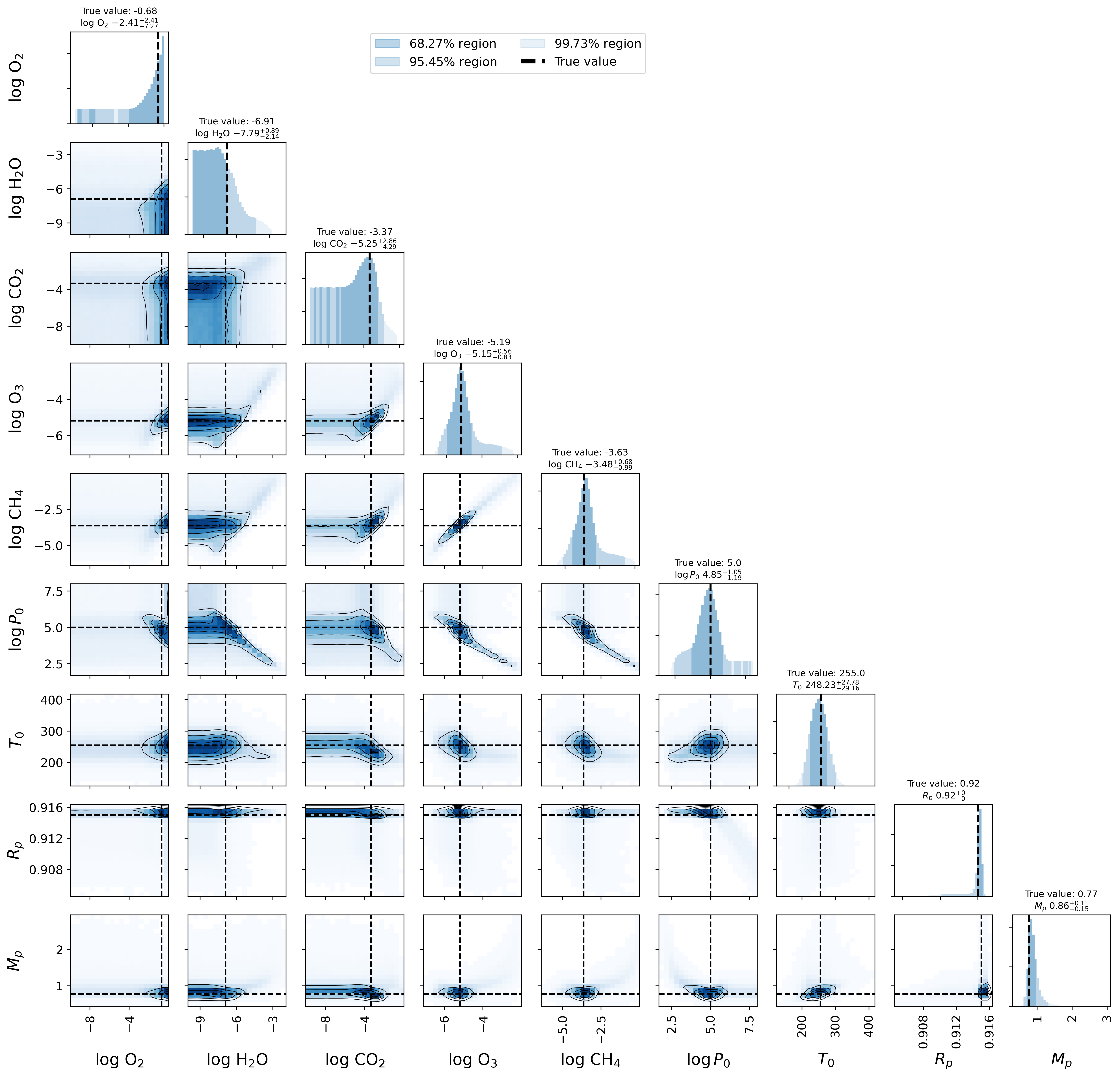}
    \caption{The full 9-parameter corner plot for the 5\,ppm simulated observational scenario presented in Section \ref{MIRI_T_S}. One-dimensional marginal posterior distributions are shown along the diagonal with associated truth values and the 16$^{\rm th}$, 50$^{\rm th}$, and 84$^{\rm th}$ percentile values indicated above each marginal distribution. The truth values for each parameter are also indicated by black dashed lines.}
    \label{ME_MIRI_5ppm}
\end{figure}

\begin{figure}[H]
    \centering
    \includegraphics[scale=0.45]{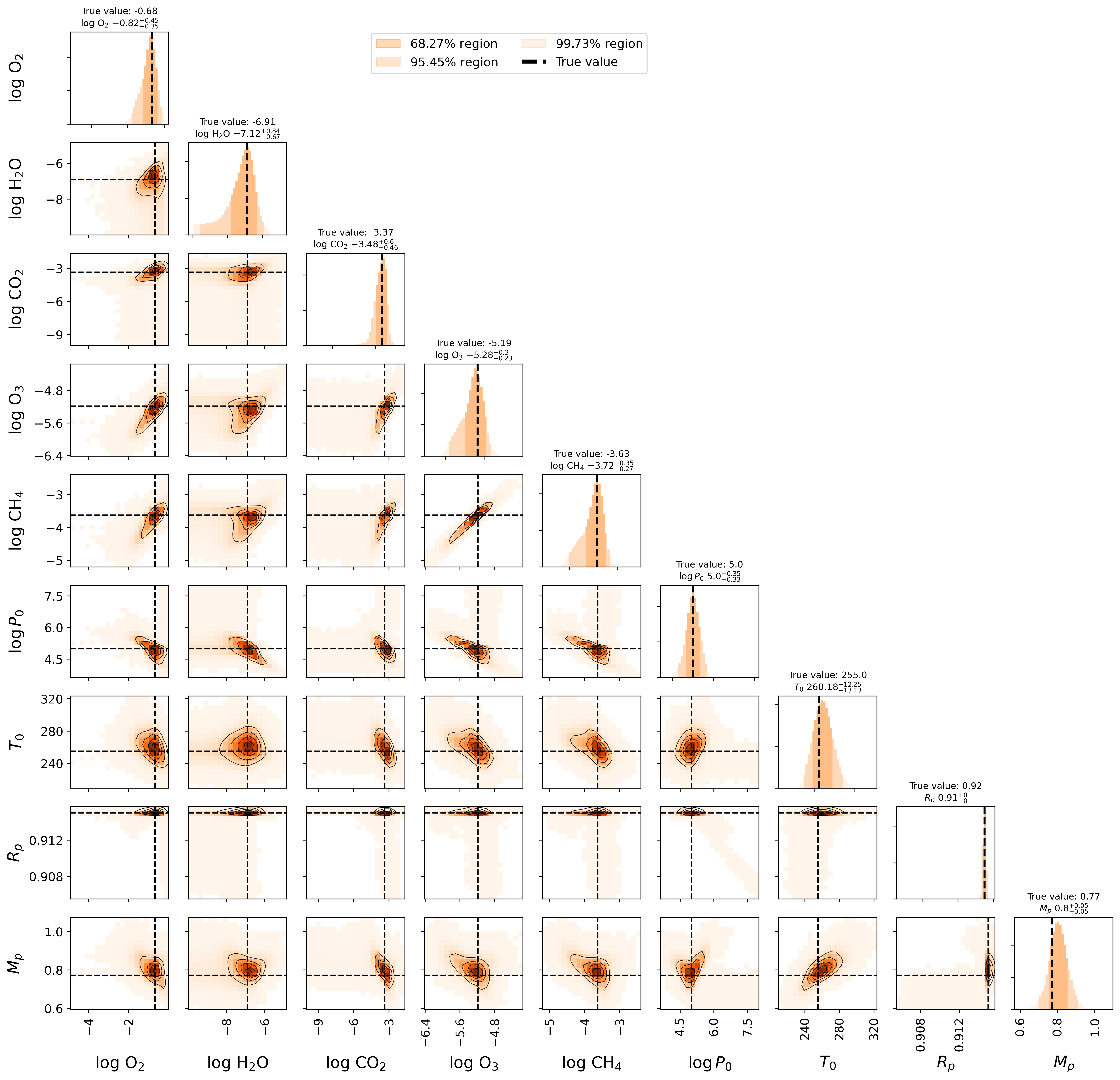}
    \caption{Same as Figure \ref{ME_MIRI_5ppm} but for the 2\,ppm scenario.}
    \label{ME_MIRI_2ppm}
\end{figure}

\begin{figure}[H]
    \centering
    \includegraphics[scale=0.45]{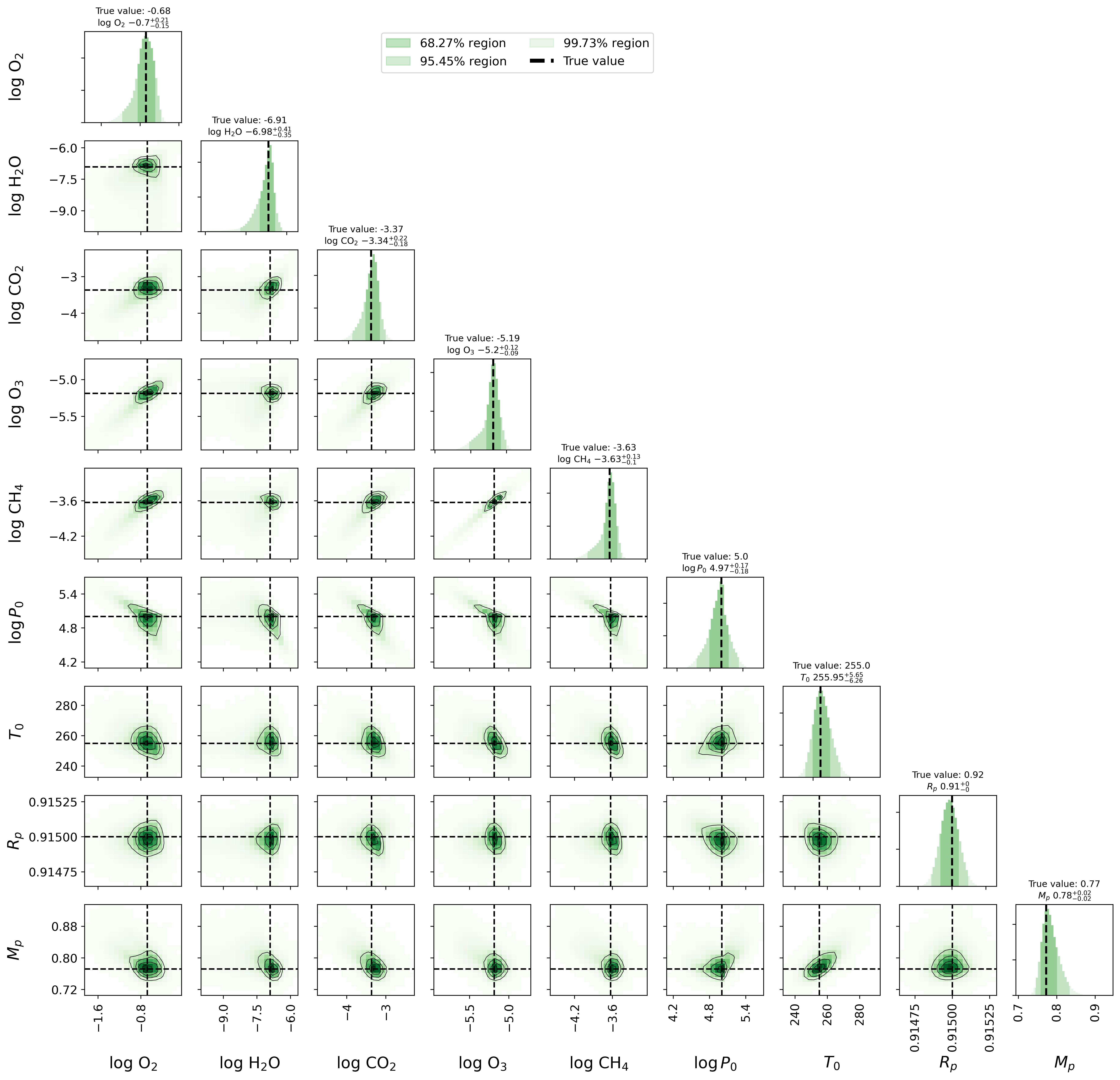}
    \caption{Same as Figure \ref{ME_MIRI_5ppm} but for the 1\,ppm scenario.}
    \label{ME_MIRI_1ppm}
\end{figure}

Figure \ref{ME_MIRI_5ppm}, Figure \ref{ME_MIRI_2ppm}, and Figure \ref{ME_MIRI_1ppm} depict the full retrieval results for the 5, 2, and 1\,ppm observational cases described in Section \ref{MIRI_T_S}. Above each 1-D posterior, the truth values, and 1-$\sigma$ confidence regions are reported.

\bibliography{references}{}
\bibliographystyle{aasjournal}



\end{document}